\documentclass[10pt]{article}
\pdfoutput=1
\usepackage{cite}
\usepackage{times}
\usepackage{amssymb,amsfonts,amsmath}
\usepackage[english]{babel}
\usepackage[pdftex]{graphicx}
\usepackage{multirow}
\usepackage{color}
\usepackage{float}
\newfloat{figtable}{ht}{lop}
\floatname{figtable}{Table}
\usepackage[labelfont=bf,labelsep=period,justification=raggedright]{caption}
\usepackage{longtable}
\usepackage{booktabs}
\usepackage{setspace}

\makeatletter
\renewcommand{\@biblabel}[1]{\quad#1.}
\makeatother

\date{}

\begin{document}

\begin{flushleft}
{\Large
\textbf{The possible role of resource requirements and academic career-choice risk on gender differences in publication rate and impact}
}
\\
Jordi Duch$^{1,2,\ast}$, 
Xiao Han T. Zeng$^{1,\ast}$, 
Marta Sales-Pardo$^{1,3}$,
Filippo Radicchi$^{3,4}$,
Shayna Otis$^{1}$,
Teresa K. Woodruff$^{5,6}$,
Lu\'is A. Nunes Amaral$^{1,4,7,\dagger}$
{\small
\\
\bf{1} Department of Chemical and Biological Engineering, Northwestern University, Evanston, IL 60208, United States of America
\\
\bf{2} Departament d'Enginyeria Inform\`atica i Matem\`atiques, Universitat Rovira i Virgili, Tarragona 43007, Spain
\\
\bf{3} Departament d'Enginyeria Qu\'imica, Universitat Rovira i Virgili, Tarragona 43007, Spain
\\
\bf{4} Howard Hughes Medical Institute, Northwestern University, Evanston, IL 60208, United States of America
\\
\bf{5} Department of Obstetrics \& Gynecology, Feinberg School of Medicine, Northwestern University, Chicago, IL 60611, United States of America
\\
\bf{6} Institute for Women's Health Research, Northwestern University, Chicago, IL 60611, United States of America
\\
\bf{7} Northwestern Institute on Complex Systems, Northwestern University, Evanston, IL 60208, United States of America
\\
}

$\ast$ These authors contributed equally to this work.

$\dagger$ E-mail: amaral@northwestern.edu
\end{flushleft}

\section*{Abstract}
Many studies demonstrate that there is still a significant gender bias, especially at higher career levels, in many areas including science, technology, engineering, and mathematics (STEM). We investigated field-dependent, gender-specific effects of the selective pressures individuals experience as they pursue a career in academia within seven STEM disciplines. We built a unique database that comprises 437,787 publications authored by 4,292 faculty members at top United States research universities. Our analyses reveal that gender differences in publication rate and impact are discipline-specific. Our results also support two hypotheses.
First, the widely-reported lower publication rates of female faculty are correlated with the amount of research resources typically needed in the discipline considered, and thus may be explained by the lower level of institutional support historically received by females. Second, in disciplines where pursuing an academic position incurs greater career risk, female faculty tend to have a greater fraction of higher impact publications than males. Our findings have significant, field-specific, policy implications for achieving diversity at the faculty level within the STEM disciplines.

\section*{Author Summary}

\section*{Introduction}
The proportion of women faculty members in many STEM fields has been steadily increasing, but at the level of associate and full professor, men continue to far outnumber women \cite{AAMC}. This is troubling because studies suggest that a lack of women in leadership positions has a negative impact on women's aspirations and advancement \cite{Carnes2008,Beaman2012} and may perpetrate gender biases \cite{Petersen2012}. Many mechanisms have been proposed to explain the gradual loss of women along the STEM academic career path \cite{Ceci2011}. For example, Carnes et al. \cite{Carnes2008} suggested that female faculty in academic medical centers experience a number of systemic and selective pressures that put them at a disadvantage at each step of their pursuit of tenure, and in achieving positions of leadership. These pressures could amount to a ``glass ceiling'' preventing women's advancement. Others have referred to the Matthew \cite{Merton1968} and Matilda \cite{Rossiter1993} effects as the cause of gender differences, that is, the greater resources awarded to men enable them to further advance their careers beyond what is possible for women. 

In contrast, to these concerns, Etzkowitz and Ranga \cite{Etzkowitz2011} recently suggested that the low number of females in academic positions within STEM disciplines should not be a cause for concern because women do not drop from STEM pursuits when they abandon academic careers but merely pursue STEM careers in other arenas. Curiously, Etzkowitz and Ranga's ``vanish box'' perspective \cite{Etzkowitz2011} does not address whether the reasons for women leaving academia do not detract from a level-playing field or whether women have the opportunity to rise to positions of prominence in non-academic careers.

To determine how and why gender may affect the professional practices and scientific production of researchers, we investigated for seven STEM fields in a quantitative manner the gender-specific and discipline-specific effects of (i) research resource requirements and (ii) relative risk in pursuing an academic career. We explicitly separated the researchers in our database along disciplinary lines in order to more carefully investigate the mechanisms potentially responsible for the observed differences. In contrast to most studies concerned with this matter, we did not conduct surveys but instead systematically analyzed the complete publication records of faculty at a large number of departments in selected research universities in the United States (Table~\ref{table1}, Fig.~\ref{fig1} and Table S\ref{tables1}--S\ref{tables7}). These data enabled us to characterize the career-long scientific production of a sizable sample of faculty from seven disciplines, and to measure statistically significant differences that would have otherwise remained hidden.

\subsection*{Data}
\begin{figure}[!ht]
\begin{center}
\includegraphics[width=\textwidth,clip=true,trim=0 0 0 0]{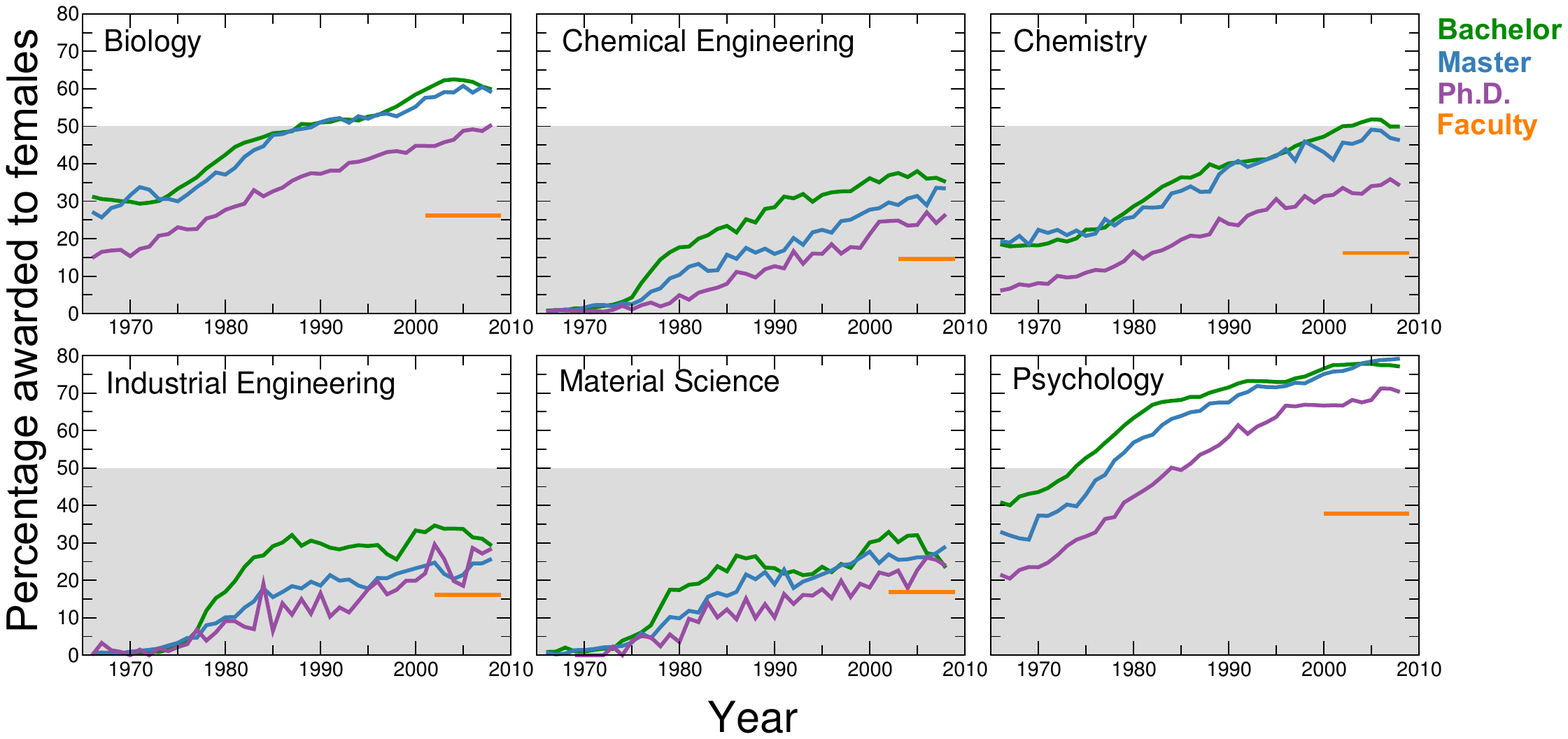}
\caption{
}
\label{fig1}
\end{center}
\end{figure}

\begin{figure}[!ht]
\begin{center}
\includegraphics[width=\textwidth,clip=true,trim=0 0 0 0]{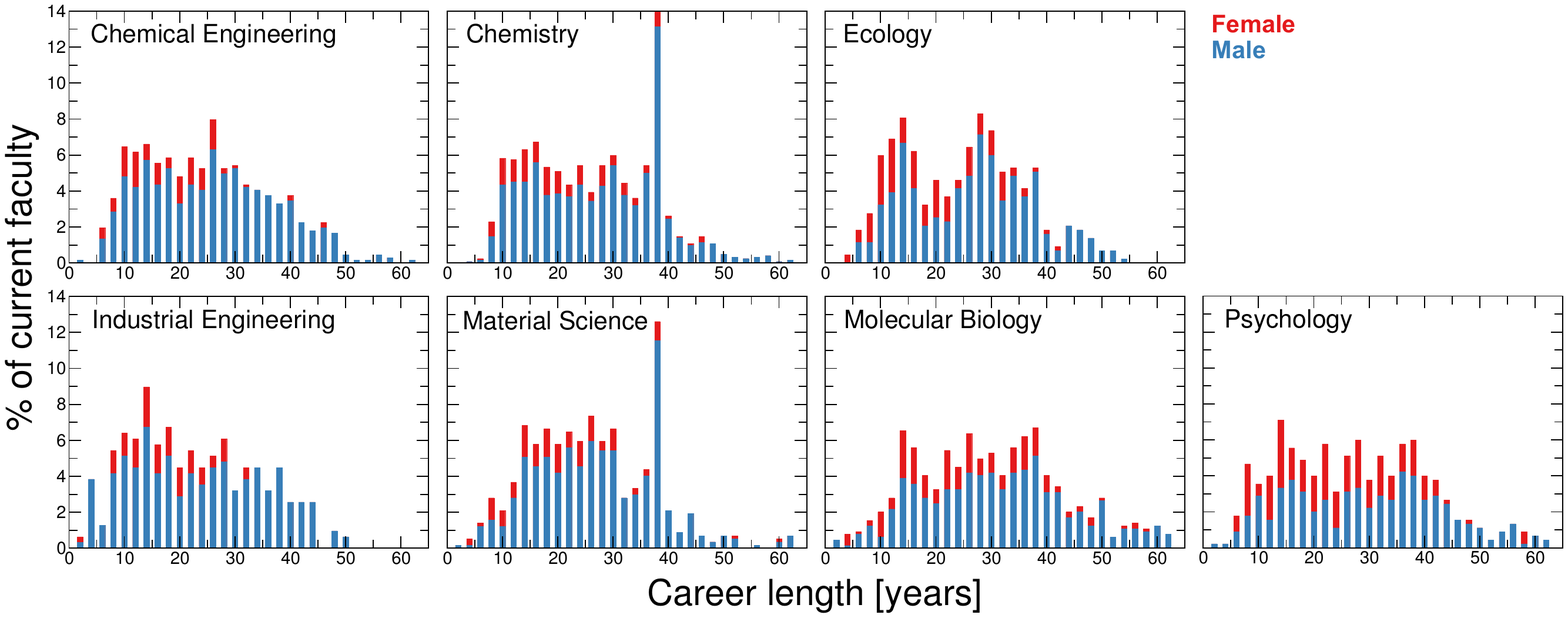}
\caption{ 
}
\vskip .1cm
\label{fig2}
\end{center}
\end{figure}

We collected data on the 2010 faculty rosters of selected top research institutions in the U.S. (see Supporting Information) in seven STEM disciplines --- chemical engineering, chemistry, ecology, industrial engineering, material science, molecular biology and psychology --- and measured scientific productivity and impact during the various phases of each faculty member's academic career \cite{USNews}. We focus on faculty at top U.S. research university departments because most high impact research produced by U.S. authors is published by authors in the top departments. We chose these disciplines for three sets of reasons. First, for all seven disciplines, women only began to join faculty rosters in a consistent manner in the 1980's, and today they still comprise a small fraction of total faculty (Fig.~\ref{fig1} and Fig.~\ref{fig2}).
Second, these disciplines cover a broad range of scientific approaches: some place greater emphasis on theoretical or computational work, whereas others focus on industrial applications or on biological systems.  Thus the requirement for institutional support --- be it lab space \cite{UPenn2001,CWRU2003,Princeton,NRC2001}, size of start-up packages\cite{UPenn2001,CWRU2003,Princeton,NRC2001}, or the ability to lead center-level projects --- required for success differs dramatically across these disciplines (Table \ref{table2}).
Third, these disciplines pose quite different relative risk profiles to individuals wishing to pursue an academic career. For example, the seven disciplines differ significantly in the prospective earnings of different career options available to Ph.D. graduates and on the time needed to achieve career stability within academia (Table \ref{table2}).

\section*{Results}

We first focus on research resource requirements. As mentioned earlier, the typical annual research expenditures per faculty member differ substantially across the seven disciplines. For example, industrial engineering faculty tend, for the most part, to train a small number of students at a time. Additionally, much of the research in industrial engineering is theoretical or computational in nature. These two characteristics suggest that, for industrial engineering, researchers do not need to compete against each other for limited resources, and institutional support may not be as important a factor in faculty productivity.

In contrast, most faculty in molecular biology conduct experimental research, and many require significant lab space and expensive specialized equipment. Moreover, faculty in molecular biology are able to compete for funding supporting the creation of large centers or the acquisition of major equipment. Thus, availability of resources, especially institutionally granted resources or institutional support for securing large grants, can be crucial components of academic success in molecular biology \cite{Princeton}. Furthermore, consistent with the Matthew effect \cite{Merton1968,Xie1998,Prpic2002,Lariviere2011}, researchers who have already received more institutional support are able to secure even more research resources.

Since historically female faculty members have received less institutional support and have had less access to research resources \cite{MIT1999,Ginther2003,Bornmann2007,Sarfaty2007,Furnham2011,Doucet2012}, these considerations prompt a question with significant policy implications: Could the differences in resource requirements lead to distinct gender-specific publication patterns across disciplines? In order to answer this question, we systematically investigated gender-specific publication rates for the seven disciplines. Even though several studies report greater publication rates by male authors \cite{Cole1984,Long1992,Xie2003,Symonds2006,Jagsi2006}, we hypothesize that only in disciplines where resource requirements are high and institutional support is vital will female faculty members typically publish fewer papers than their male peers. Thus, we predict that gender differences in publication rate in disciplines such as industrial engineering are going to be quite low. In contrast, we predict that gender differences in publication rate are going to be very significant in molecular biology and similar disciplines.

We define the publication rate of a faculty member $s$ years into her/his career as the number of scientific articles published by the individual $s$ years after her/his first publication. We cannot simply compare the raw publication numbers per year, because these numbers depend strongly on publication year $y$ and career stage $s$ (Fig.~\ref{fig3}). Let $n_i(y)$ denote the number of publications published by author $i$ from discipline $j$ in year $y$, and let $N_j(y)$ be the total number of authors that have started their careers no later than year $y$. We calculate author $i$'s z-score (standard score) in year $y$ as 
\begin{equation}
z_i(y)=\frac{n_i(y)-\mu_j(y)}{\sigma_j (y)}\;\;,
\label{eq1}
\end{equation}
where $\mu_j(y)$ is the average number of publications per author from discipline $j$ published in year $y$
\begin{equation}
\mu_j(y)=\frac{1}{N_j(y)}\sum\limits_k \, n_k(y)\;\;,
\label{eq2}
\end{equation}
and $\sigma_j(y)$ is the standard deviation of the number of papers per author published in year $y$
\begin{equation}
\sigma_j(y)=\sqrt{\frac{1}{N_j(y)}\sum\limits_k \left[n_k(y) - \mu_j(y)\right]^2} \;\;.
\label{eq3}
\end{equation}
In order to account for the effect of career stage, we consider $z^c_i(s)$, which is the z-score of author $i$ as a function of the career stage $s=y-y_i$, where $y_i$ is the year of the first publication of author $i$ (Fig.~\ref{fig4}, S\ref{figs1}). Please note that by considering the z-score we are not making any assumption about normality of $n_i(y)$, but merely making the results easier to compare across disciplines and time periods.

Our analysis fully confirms our hypothesis (Fig.~\ref{fig3}--\ref{fig5}). As predicted, for disciplines where research expenditures are high, such as molecular biology, we found that females consistently publish at a rate significantly lower than males, whereas for industrial engineering we do not observe a significant difference between genders. More importantly, as shown in Fig.~\ref{fig5}, we found that the gender difference in publication rate, measured as the average z-score of females, has a significant negative correlation with magnitude of typical research expenditures. Our results thus support the hypothesis that gender differences in institutional support have had a crucial effect on the publication rates of females. 

It is important to point out that in our analysis we did not consider human and social capital such as collaboration level and leadership position, which may also have critical roles for a productive career \cite{Petersen2012}, as research resources. Whether and how the gender difference in the ability to acquire these resources harder to quantify affects career productivity is a matter worth of further investigation.

We next investigated gender-specific and discipline-specific effects of career relative risk profile of an academic career on publication patterns. The risk to pursue a faculty position after obtaining a Ph.D. varies across disciplines. A graduate student considering an academic career in chemistry faces a small risk if unsuccessful. Within
about six years from publication of their first paper, successful individuals will move into independent positions (Fig.~\ref{fig6}, S\ref{figs2}--\ref{figs3} Table \ref{table2} and Methods). Doctoral degree holders in chemistry unable or uninterested in obtaining academic positions can chose from among a number of high-paying careers in industry and government.

In contrast, an individual considering an academic career in ecology faces a much more uncertain future. Instead of waiting six years post publication of the first paper to learn whether it will be possible to secure a faculty position, an ecologist has to wait an average of {\it eight} years (Fig.~\ref{fig6}, S\ref{figs2}--\ref{figs3}, Table \ref{table2}). Perhaps even more challenging, doctoral degree holders in ecology who are not able or not interested in obtaining academic positions may have to settle for jobs that do not pay a significant premium over academic positions.

These observations raise a critical question: Could the different risk profiles of STEM disciplines lead to distinct gender-specific selective pressures? Because pursuing an academic career is a risky undertaking and because propensity towards risk-taking \cite{Byrnes1999,Harris2006}, self-motivation towards career development \cite{Cech2010}, social expectations \cite{Ceci2009}, perception of gender stereotypes \cite{Heilman2001} and biological constraints \cite{Ceci2009,Mason2002,Mason2004} are different for females and males, we surmise that a female will choose to pursue an academic career in ``high-risk" disciplines, such as ecology, only if she is so highly qualified that she will be quite confident of success. This biased self-selection for outstanding individuals among females likely happens prior to embarking on an academic career \cite{Kaminski2012}, leading to females' advantage in career performance that would be magnified in later stages of career due to the Matthew effect \cite{Petersen2011}. In contrast, because of the low risk profile of chemistry, we expect that female faculty members in chemistry will incur no extra burden when compared to their male colleagues. It is worth mentioning that an alternative hypothesis is that high career risk induces selection for individuals with greater propensity to risk-taking among females. However, this is consistent with our hypothesis, since risk-taking might be a necessary ingredient, among other intellectual abilities, towards success, and individuals may augment their competence through risk-taking. Therefore, females who enter disciplines with high career risks may be not only risk-takers but in fact also highly qualified.

We further hypothesize that the higher qualification of females in high-risk disciplines will become apparent through higher impact per publication. In order to uncover gender differences in publication impact, we studied a commonly used metric of academic performance, the $h$-index \cite{Hirsch2005}. We studied the $h$-index instead of the total number or average number of citations because the distributions of these numbers can be dramatically biased by a single highly-cited publication \cite{Bornmann2007a}. The $h$-index avoids this bias by identifying the number of publications of an author that have at least that number of citations. Moreover, because the $h$-index was introduced after the time period considered for the data, it will not be affected by behaviors of the authors aimed at deliberately increasing their $h$-indices. 

An identified weakness of the $h$-index is its dependence on the number of publications. In order to compare the publication impact of authors with different number of publications, we determined the dependence of the $h$-index on the number of publications for the faculty cohorts in the seven disciplines considered. We found that for these seven disciplines the $h$-index grows with the number of publications as a power law \cite{VanRaan2006},
\begin{equation}
h = k\, n^{\alpha}\,,
\label{eq4}
\end{equation}
where $n$ is the number of publications (Fig.~\ref{fig7} and Methods). For $\alpha = 1$, the $h$-index would grow linearly with number of publications. Importantly, since we find $\alpha \approx 0.6$, one cannot explain the observed values of $\alpha$ through self-citations alone (Methods).

We next measured the deviations of \emph{h}-indices from the trend predicted by
Eq.~(\ref{eq4}) for individual faculty members to obtain the z-scores (standard score) of their publication impact (Fig.~\ref{fig8}). Let $h_i$ denote the \emph{h}-index of author $i$, and $n_i$ her/his total number of publication. The z-score of \emph{h}-index of author $i$ is
\begin{equation}
\zeta_i=\frac{h_i-kn_i^\alpha}{\sqrt{kn_i^\alpha}}\;\;.
\label{eq5}
\end{equation}
We then calculated the average z-scores of this publication adjusted $h$-index of females (Fig.~\ref{fig8}, S\ref{fig4} and Methods). Our analysis unambiguously shows that for all ranges of number of publications, female faculty members in ecology published research with higher impact than their male counterparts, whereas for faculty in chemistry we found no significant gender-specific differences in impact. 

The data in Fig.~\ref{fig8} suggest that the difference $D$ in publication impact may be an increasing function of the discipline-specific risk profile $R$ associated with an academic career. That is,
\begin{equation}
D=a_0+a_1R+\cdots\;\;.
\label{eq6}
\end{equation}
While we lack a theory for the true definition of career risk, $R$, it is plausible that it will be a function of factors such as the time $T$ to reach career independence, the fraction $A$ of Ph.D. graduates that go on to careers in academia, and the reciprocal of the salary premium of non-academic careers (Table \ref{table2}, \cite{NSFStat}), which we define as
\begin{equation}
P=\frac{S_\mathrm{academic}}{S_\mathrm{non-academic}-S_\mathrm{academic}}\;\;. 
\label{eq7}
\end{equation}
Even though we do not know its functional form, we can expand $R$ as a multivariate polynomial,
\begin{equation}
R(T,P,A)=b_0+b_1T+b_2P+b_3A+b_4TP+b_5PA+b_6TA+b_7T^2+\cdots\;\;,
\label{eq8}
\end{equation}
and it follows that we can expand $D$ as
\begin{equation}
D(T,P,A)=c_0+c_1T+c_2P+c_3A+c_4TP+c_5PA+c_6TA+c_7T^2+\cdots\;\;.
\label{eq9}
\end{equation}
Because we only have $7$ data points, we must fit our data to combinations of at most $2$ terms in the expansion. Ordinary least squares regression indicates that the difference in publication impact across the seven disciplines is positively correlated with several combinations of the factors in Eq.~(\ref{eq9}), thus confirming the existence of the relative risk associated with academic careers and its gender-specific role on publication impact (Table ~\ref{table3}). In Fig.~\ref{fig9} we show the correlation between the gender difference in publication impact and the academic career risk, quantified as
\begin{equation}
R=d_0+d_1P+d_2TA\;\;.
\label{eq10}
\end{equation}
This model suggests that in disciplines where there are few non-academic career options available and the time to reach career independence is long, and where it is difficulty to recover salary loss due to unsuccessful academic career, pursuing an academic position is highly risky.

\section*{Discussion}
Our study reveals the possible contribution of perceived risk and resource allocation to the under-representation of women in STEM academic careers. Our results are not by themselves an empirical validation of the causal relationship between publication rate and resource requirements, and between publication impact and career risk, since we cannot conduct controlled experiments or account for other factors that could play a role in the measured outcome. However, the hypothesis that there is a causal relationship between gender differences in resource allocation and the reported gender differences in publication rates is plausible and well supported by our empirical observations, as is the hypothesis that there is a causal relationship between the relative risk associated with academic careers and the gender differences in publication impact.

The issues we identify here, together with the known socialization concerns surrounding work-life balance, may have created a ``tipping point" that explains the nearly intractable problem of retaining women within STEM disciplines. It is equally important to think about the role these previously unrecognized risk factors may contribute to the number of under-represented minorities in the STEM pipeline.  It is not possible to address this point using the methods we describe here, but there may be opportunity and new impetus to develop novel tools that can provide a more sophisticated insight into why some groups of people are not well represented in scientific subspecialties. More intriguingly, we wonder how the perceived or real risks associated with resource infrastructure and future opportunities can be translated into other fields (business, politics, the legal profession) where there is a paucity of women and minorities in the upper career rungs. Most importantly, now that these factors have been identified, it should be possible to create policies that provide better opportunities for all individuals with an aptitude for science, and perhaps in all kinds of careers, to ensure that our work force is diverse and can gain from the insights of all contributing members.

\section*{Methods}
\paragraph{Data acquisition.} 
We obtained complete faculty rosters as of June,
2010 for several top research universities in the U.S. in the
disciplines of chemical engineering, chemistry, ecology, industrial engineering, material science, molecular biology and psychology (see Table S\ref{tables1}--\ref{tables7} for a complete list
of institutions and departments that were included in our
analysis). We considered all active faculty members, including
tenure-track and research faculty, but excluded emeritus
professors. For each faculty member, we collected the following data:
gender, year of Ph.D. (if available), current and past positions, a
list of publications published by the end of 2010 and indexed in Thomson Reuters Web of Science (WoS), and the number of citations for these publications as of June, 2011. To obtain a
reliable list of publications for each investigator from the WoS, we
designed a supervised disambiguation protocol. Our protocol uses
biographic information for an investigator to build and refine a
query that retrieves the entire list of publications from the WoS. For example:
\begin{enumerate}
\item Select last name and set of initials that the investigator could
  potentially use to sign her papers. For instance, David A. Tirrell has two potential WoS names, ``Tirrell D" and ``Tirrell DA."
\item Set the year of publication range from four years before the Ph.D.
  date until the data acquisition time. If the Ph.D. year is not
  available, estimate the Ph.D. year from the list of publications
  listed in the investigator's personal web page or from the date of
  hire. For David A. Tirrell, our protocol returns the publications from 1974 on for ``Tirrell D" and ``Tirrell DA" (1974 = 1978 - 4 and 1978 is the year Professor Tirrell was awarded a Ph.D.).
\item When current and previous positions are available, constrain the
  query to retrieve publications that include one of those
  institutions as one of the author's address.
\end{enumerate}
The disambiguation protocol downloads all types of publications of the authors. In the analysis we included articles, conference proceedings and reviews. At each step, we obtained the number of publications assigned to a
particular author and checked for anomalies using a number of data
features, the most important of which were:
\begin{enumerate}
\item The total number of publications is consistent with the current
  position of the investigator, the number of years doing research,
  and the type of research.
\item The number of publications in each year does not deviate ``significantly" from the average of the surrounding years.
\item Journal titles of the publications are within the investigator's
  field of expertise.
\end{enumerate}
Our disambiguation protocol allows us to introduce different names or initials for each scientist. For example, for females, for whom there is evidence in the list of publications of their CVs that they change their family name after marriage, we include both names in the query. Note that the errors in the publication list introduced by name changes is small \cite{Radicchi2009}.
To estimate the percentage of false positives in the publications assigned to an author, we randomly sampled about one hundred authors in our database who had an updated list of publications on their personal websites. We then manually checked these lists against the results we obtained from the WoS. We estimated that, using our disambiguation protocol, the percentage of false positives in the publications assigned to an author is less than $2\%$.

\paragraph{Number of publications and $h$-index distributions.}
For the analysis of the $h$-index and the number of publications, we
considered only papers published by December 31$^{st}$, 2000. In order to have a
reliable measure of the $h$-index, we need to consider papers which
have accrued a number of citations that truly reflects the impact of
that research. Based on prior studies \cite{Stringer2008}, we set
ten years as the threshold for papers to have
accumulated their ``ultimate'' number of citations.

\paragraph{The value of $\alpha$ due to self-citations.}
Assume that an author with $n$ publications makes $r$
  self-citations in each of her/his publications. The total number of
  self-citations is thus $rn$. In order to maximize his/her
  \emph{h}-index, the author will distribute his self-citations
  homogeneously among $h$ of his own publications. Thus, the average
  number of citations per publication is $rn/h = h$, yielding $h^2 = rn$
  or $h \propto n^{1/2}$. That is, $\alpha = 0.5$.

\paragraph{Fitting the $h\propto n^{\alpha}$ relationship.}
We surmise that given the number of publications $n$, the $h$-index
$h$ is a random variable obeying the Poisson distribution:
\begin{equation}
p(h|n) =  e^{- \lambda (n)}\lambda (n)^h/ h! \,,
\end{equation}
with mean $\lambda (n)= kn^{\alpha}$. The likelihood of the data given
this model is then:
\begin{equation}
\mathcal{L} = \prod \limits_i p(h_i|n_i) \,,
\end{equation}
where the product runs over all pairs $(h_i, n_i)$ in real data. The
best estimates of $k$ and $\alpha$ are those that maximize
$\mathcal{L}$. The estimates yield good fits to the data (see Fig.~\ref{fig7}).

\paragraph{Fitting the transition to independence.}
We fitted the data in Fig.~\ref{fig6} to the generalized logistic
function,
\begin{equation}
Y(t)=A+\frac{K-A}{1+\textrm{e}^{-B(t-M)}} \,,
\end{equation}
where $A$ is the lower asymptote, $K$ is the upper asymptote, $B$ is
the growth rate, and $M$ is the time of maximum growth. We provide the
values of the fitting parameters for all data sets in Tables S\ref{tables11} --S\ref{tables17}. We use $M$ as a proxy for the time for transition to professional independence.

\paragraph{Statistical significance of linear correlations.}
The p-values of the linear correlations in Figure 5 and Figure 9 are obtained using two statistical tests, the permutation test and Student's t-test. Since the Student's t-test is well known, we describe here only the permutation test. Suppose that we have $N$ data points on the two dimensional plane. We consider all the $N!$ permutations of the $x$ (or $y$) values of the data points, and calculate the correlation coefficient for each of the permutation, which will yield $N!$ correlation coefficients, $R_1$, $R_2$, $\cdots$, $R_{N!}$. We then calculate the probability that these coefficients are larger than or equal to the correlation coefficient of the original data set $R_o$, $P(R_i\geq R_o)$. This probability is the p-value given by the permutation test.

\section*{Acknowledgments}
We thank R. Guimer\`a, S. Mukherjee, R. D. Malmgren, P. McMullen, M. J. Stringer, and James A. Evans for comments and suggestions. We thank S. C. Tobin for editorial assistance.  L. A. N. Amaral gratefully acknowledges the support of NSF awards SBE 0624318 and 0830388, and ThomsonReuters for access to the WoS data. J. Duch and M. Sales-Pardo's work have been partially supported by the Spanish DGICYT under project FIS2010-18639.

\bibliography{Manuscript_PLoSONE}

\clearpage
\newpage
\section*{Figure Legends}
\paragraph*{Figure 1}
\textbf{The leaking pipeline.}
Percentage awarded to females of the total number of bachelor (green lines), master (blue lines) and doctoral (purple lines) degrees in the period 1966--2008. We obtained these data from~\cite{NSFStat}. We also show the percentage of female faculty in our datasets (orange lines). We could not obtain separate data for molecular biology, so we show the data for biology instead. The grey shaded areas indicate values lower than 50\%. The gender ratio of the faculty members given by our data is close to that reported elsewhere. For example, in our data, the percentage of female faculty members in chemistry is 16.2\%, and according to the report of Chemical \& Engineering News, this percentage is 17\%.

\paragraph*{Figure 2}
\textbf{Career lengths of faculty members.}
Career length distribution of female (red) and male (blue) current
faculty members for a selected set of
U.S. universities (Table \ref{table1}). Data is binned
into two year intervals. Currently, females hold about 16\% of faculty positions in
chemistry and in material science departments, and about 25\% of faculty positions in molecular biology departments.

\paragraph*{Figure 3}
\textbf{Average number of annual publications per author.}
Average number of publications authored by females (red) and males (blue) as a function of time. Data is smoothed using moving averaging over a $3$-year time window. Note the increasing trend in all disciplines. Because of these trends, we must account for the different starting years and career stages of authors when comparing publication rates. 

\paragraph*{Figure 4}
\textbf{Gender difference in publication rate.}
Average z-score of number of publications for females (red) and males (blue) as a function of career stage. Shaded areas indicate the standard errors. See Fig.~S\ref{figs1} for the statistical significance of the gender difference in publication rate.

\paragraph*{Figure 5}
\textbf{Lower publication rates of female faculty is correlated with higher requirements for research resources.}
Effects of the magnitude of the resource requirements on the difference in publication rates between genders. Ecology is not included as we could not obtain data for resource requirements. The difference in publication rates is measured by the average z-score of number of publications by females in each year, and the error bars indicate the standard errors. The resource requirements is defined as the average annual research expenditure per principal investigator in the departments studied (Table \ref{table2},\cite{NSFStat}). The trend line (black dashed line) indicates a negative correlation (coefficient of determination $R^2=0.72, ~p<0.04$). These data suggest that higher resource requirements lead to greater differences in the publication rates between females and their male peers.

\paragraph*{Figure 6}
\textbf{Time to career independence.}
Fraction of publications in which a faculty member is the last author (purple diamonds) and the fraction of publications in which a faculty member is the first author (green squares). In many disciplines, the senior author of a study is listed last. Looking at the change in the fraction of times a faculty member in our dataset is a first or last author can thus be used as a proxy for change in seniority-level of an individual in these disciplines. We order publications, excluding single-author publications, by years after first publication and aggregate within each discipline. We fit the data to generalized logistic functions (green/purple lines) and define career independence (grey shaded areas) as the mid-point of the logistic function for fraction of last-author publications (Methods, Table S\ref{tables11}--\ref{tables17}). While we do not observe gender effects (Figs. S\ref{figs2}, S\ref{figs3}), we do observe differences between fields.

\paragraph*{Figure 7}
\textbf{Relation between impact and number of publications.}
Dependence of the \emph{h}-index on number of publications for faculty with at least 30 publications that are at least 10 years old (Table~S\ref{tables18}). We consider only publications at least 10 years old in order to ensure that they all have accrued close to their ultimate number of citations \cite{Stringer2008}. Blue and red dots show values for individual male and female faculty members in our cohorts. The solid black line shows a maximum likelihood power-law fit to $h(n)$ under the assumption of Poissonian fluctuations (Methods). Shaded areas indicate one standard deviation (dark grey areas) and two standard deviations (light grey areas) from the mean.

\paragraph*{Figure 8}
\textbf{Comparison of publication impact for authors with different numbers of publications.}
The z-score of the $h$-index as a function of number of publications. We use the mean and standard deviation obtained from the parameters in the model to determine the z-scores.

\paragraph*{Figure 9}
\textbf{Higher publication impact of female faculty is correlated with higher relative risk of academic career choice.}
Risk in academic career choice and difference in publication impact. We quantify the risk of academic career choice according to Eq.~(\ref{eq10}). We show results for two alternative measures of difference in publication impact. In ({\bf A}), we defined the gender difference in publication impact as the average {\emph h}-index z-scores of females. The error bars indicate standard errors. See Fig.~S\ref{figs4} for the statistical significance of the gender difference in publication impact. The trend line (black dashed line) indicates a significant positive correlation (coefficient of determination $R^2=0.96, ~p=0.001$). In ({\bf B}), we defined the gender difference in publication impact as the probability that female authors have larger \emph{h}-index z-scores than male authors, as depicted in Fig. S\ref{figs4}. The trend line (black dashed line) indicates a significant positive correlation (coefficient of determination $R^2=0.86, ~p=0.005$). Note that the values of the risk of academic career choice in ({\bf A}) and ({\bf B}) are different for each discipline because the coefficients in the linear regression are different. The data suggest that in disciplines where it is risky to pursue an academic career, female faculty have publications with higher impact than male faculty. 

\clearpage
\newpage

\section*{Tables}
\begin{table}[h]
\caption{{\bf Female and male cohorts in study.}}
\begin{center}
\begin{tabular}{lcrrrrr}
\multirow{2}{*}{{\bf Discipline}} & \multirow{2}{*}{{\bf Departments}} & \multicolumn{2}{c}{{\bf Female}} & & \multicolumn{2}{c}{{\bf Male}}\\\cline{3-4}\cline{6-7}
& & Authors & Publications && Authors & Publications\\ \midrule
Chemical Engineering & 31 & 98 & 6,392 && 567 & 66,328\\
Chemistry & 35 & 198 & 13,790 && 1,020 & 137,723\\
Ecology & 15 & 106 & 3,976 && 328 & 22,425\\
Industrial Engineering & 15 & 51 & 1,498 && 261 & 11,509\\
Material Science & 26 & 98 & 9,538 && 473 & 75,373\\
Molecular Biology & 11 & 168 & 9,882 && 474 & 51,234\\
Psychology & 10 & 171 & 7,143 && 279 & 20,976\\\midrule
{\bf Total} & &890 & 52,219 && 3,402 & 385,568\\
\end{tabular}
\label{table1}
\end{center}
\end{table}

\clearpage
\newpage
\begin{table}[h]
\begin{small}
\caption{{\bf Requirement for research resources and risk of academic career choice.} $T$, time to reach career independence. $P$, reciprocal of the salary premium of non-academic careers. $A$, ratio of Ph.D. graduates pursuing an academic position. The data of research expenditures are obtained from http://www.nsf.gov/statistics/nsf11313/. The salary data are obtained from http://www.nsf.gov/statistics/nsf09317/. We use the salary data of zoology instead of ecology because we could not obtain data for ecology. The data of career choice of Ph.D. graduates are obtained from http://www.nsf.gov/statistics/nsf03310/.}
\begin{center}
\begin{tabular}{lcrcccc}
\multirow{3}{*}{{\bf Discipline}} & \bf Avg. annual & \multicolumn{2}{c}{{\bf Median of salary [K\$]}} & {\bf Salary premium of} & {\bf Time to} & \bf Frac. of graduates \\\cline{3-4}
& \bf expediture &&& \bf non-academic & \bf career & \bf pursuing\\
& \bf per PI [M\$] & Acad. & Non-acad. & \bf careers, $P^{-1}$ & \bf  independence, $T$ & \bf  acad. careers, $A$ \\ \midrule
Chemical Engineering & 0.490 & 77.2 & 107.2 & 0.39 & 5.4 & 0.21\\
Chemistry & 0.515 & 64.6 & 104.2 & 0.61 & 6.2 & 0.32\\
Ecology & --- & 69.0 & 95.0 & 0.38 & 8.2 & 0.71\\
Industrial Engineering & 0.094 & 74.0 & 104.0 & 0.41 & 6.1 & 0.56\\
Material Science & 0.612 & 74.5 & 102.5 & 0.38 & 6.6 & 0.20\\
Molecular Biology & 1.897 & 62.2 & 100.9 & 0.62 & 7.3 & 0.57\\
Psychology & 0.256 & 65.6 & 96.2 & 0.47 & 8.2 & 0.42\\
\end{tabular}
\label{table2}
\end{center}
\end{small}
\end{table}

\clearpage
\newpage
\begin{table}[h]
\begin{small}
\caption{{\bf Linear models predicting the gender difference in publication impact.} The gender difference in publication impact is defined as the average {\emph h}-index z-scores of females. $T$, time to reach career independence. $P$, reciprocal of salary premium of non-academic careers. $A$, ratio of Ph.D. graduates pursuing an academic position. $ ^{\dagger} p<0.10,\; ^{*}p<0.05,\; ^{**}p<0.01$. The $p$-values indicated below were obtained with the permutation test, but using Student's t-test yields similar results.}
\begin{center}
\begin{tabular}{cccccccc}
Intercept & $P$ & $A$ & $TP$ & $PA$ & $TA$ & Adj. $R^2$ & $p$-value\\\midrule
~~~~$-0.96^{**}$ & ~~~~$0.39^{**}$ &&&& $0.10^{**}$ & $0.94$ & $0.001$\\
$[-1.32,~-0.61]$ & $[0.25,~0.54]$ &&&& $[0.06,~0.14]$ &&\\
~~~~$-1.03^{**}$ & ~~~~$0.40^{**}$ & ~~$0.79^*$ &&&& $0.88$ & $0.001$\\
$[-1.57,~-0.50]$ & $[0.19,~0.61]$ & $[0.29,~1.28]$ &&&&&\\
~~$-0.68^{*}$ & & $0.37$ & ~~$0.05^{*}$ &&& $0.79$ & $0.007$\\
$[-1.12,~-0.17]$ & & $[-0.32,~1.06]$ & $[0.01,~0.08]$ &&&&\\
~~$-0.62^{*}$ & && ~~~~$0.05^{**}$ &&& $0.74$ &$0.007$\\
$[-1.14,~-0.10]$ & && $[0.02~0.09]$ &&&&\\
%
%
$-0.17$ & &&& ~~$0.40^*$ && $0.65$ &$0.02$\\
$[-0.48,~0.14]$ & &&& $[0.11,~0.69]$ &&&\\
$-0.64$ & $0.38$ &&&&& $0.42$ &$0.01$\\
$[-1.59,~0.32]$ & $[-0.04,~0.79]$ &&&&&&\\
%
%

\end{tabular}
\label{table3}
\end{center}
\end{small}
\end{table}

\newpage
\section*{Figures}

\clearpage

\clearpage
\begin{figure}[!ht]
\begin{center}
\vskip .7cm
\includegraphics[width=\textwidth,clip=true,trim=0 0 0 0]{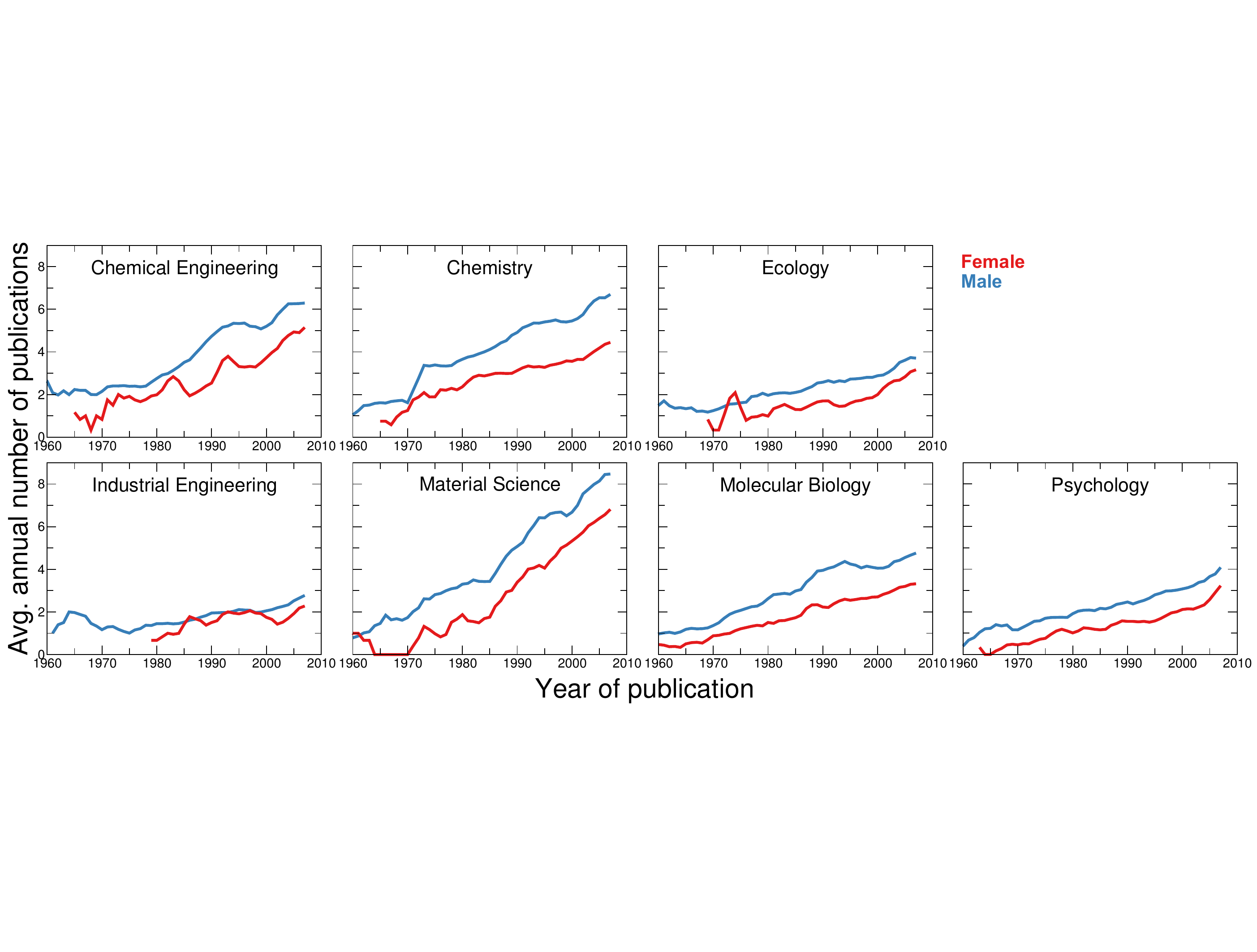}
\caption{
}
\label{fig3}
\end{center}
\end{figure}

\clearpage
\begin{figure}[!ht]
\begin{center}
\vskip .7cm
\includegraphics[width=\textwidth,clip=true,trim=0 0 0 0]{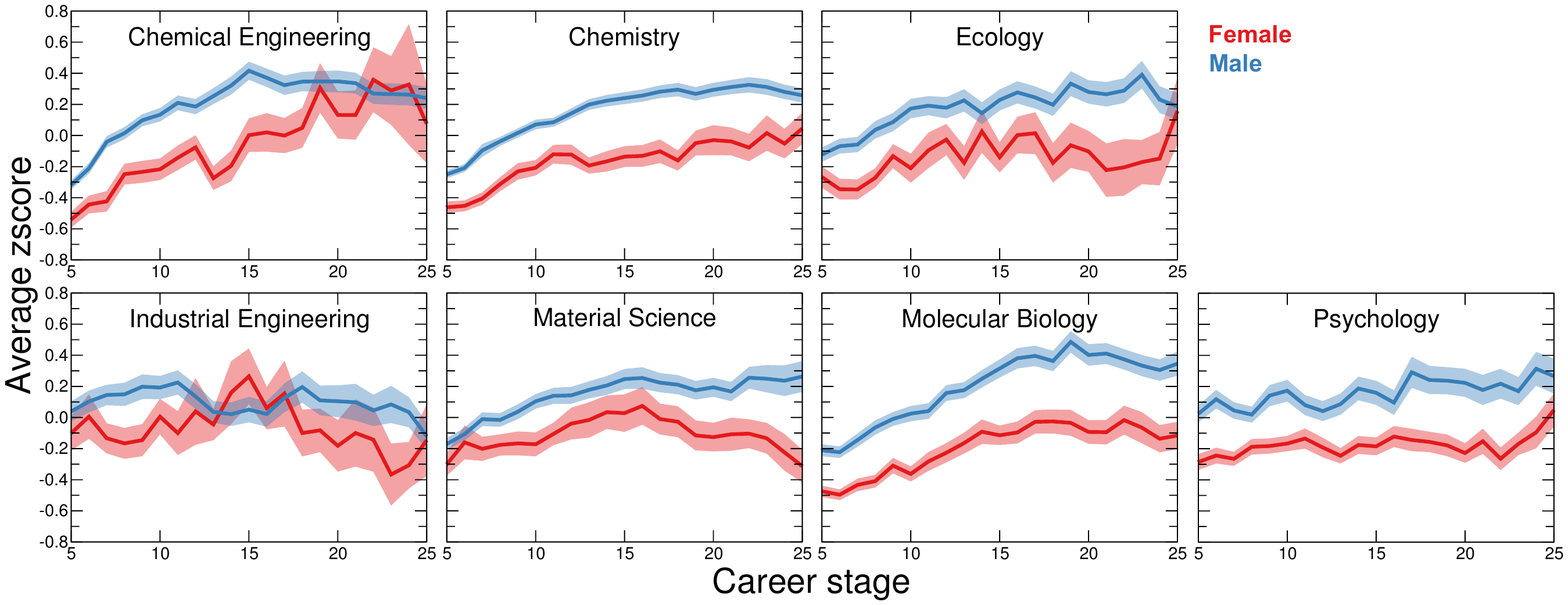}
\caption{ 
}
\label{fig4}
\end{center}
\end{figure}

\clearpage
\begin{figure}[!ht]
\begin{center}
\vskip .7cm
\includegraphics[width=\textwidth,clip=true,trim=0 0 0 0]{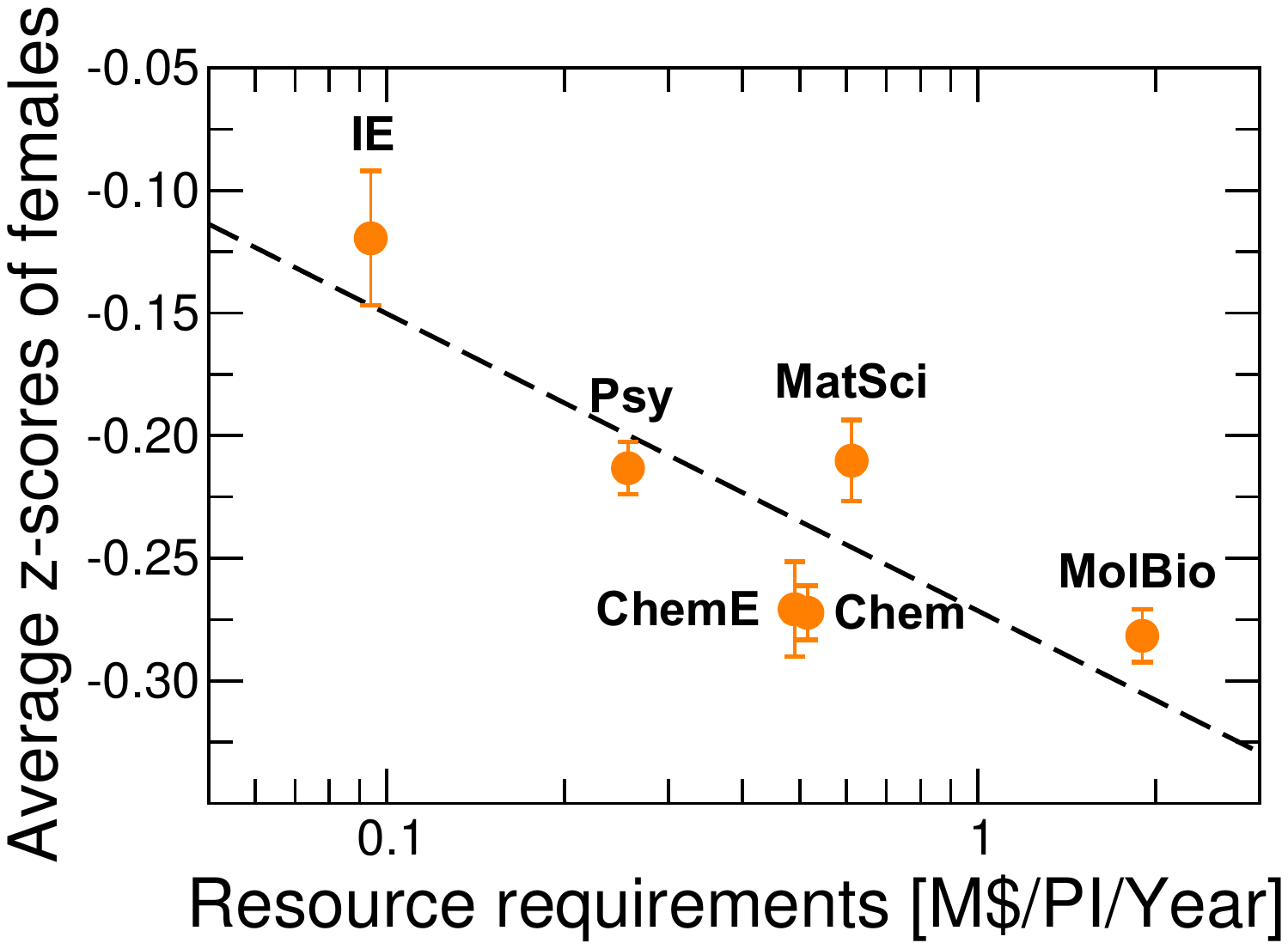}
\caption{ 
}
\label{fig5}
\end{center}
\end{figure}

\clearpage
\begin{figure}[!ht]
\begin{center}
\vskip .7cm
\includegraphics[width=\textwidth,clip=true,trim=0 0 0 0]{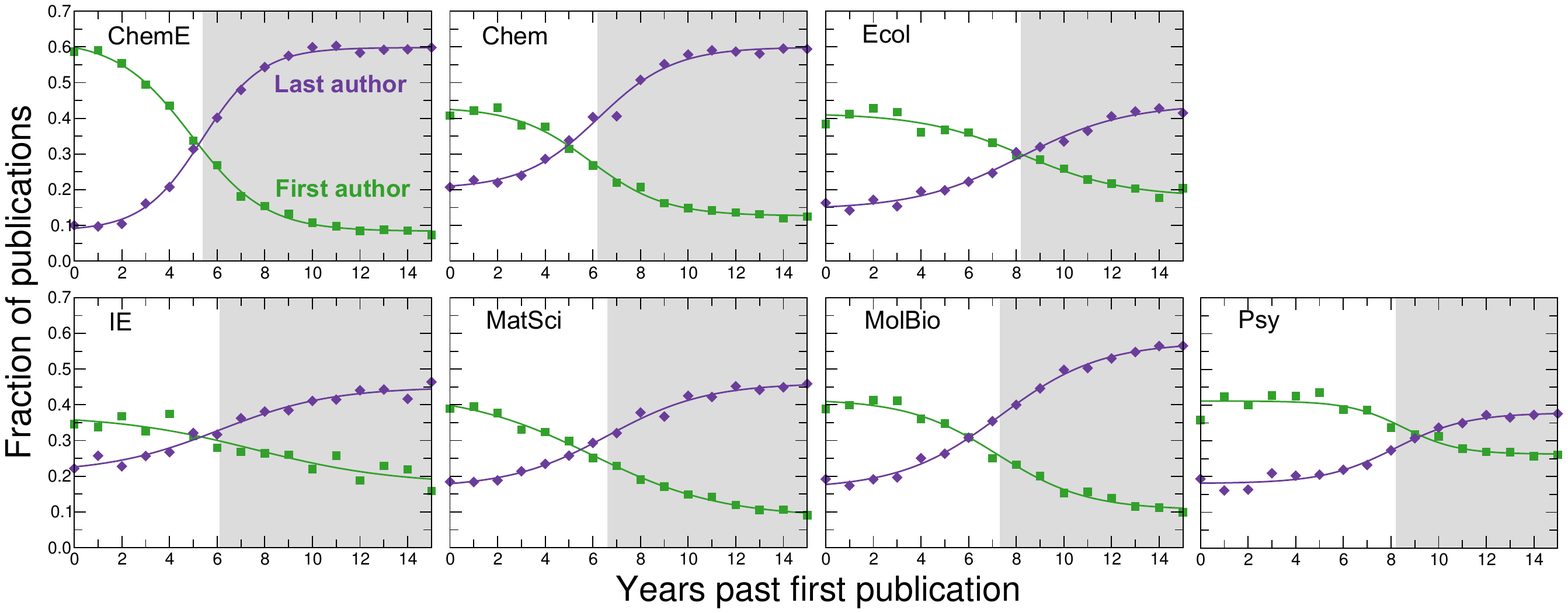}
\caption{ 
}
\label{fig6}
\end{center}
\end{figure}

\clearpage
\begin{figure}[!ht]
\begin{center}
\vskip .7cm
\includegraphics[width=\textwidth,clip=true,trim=0 0 0 0]{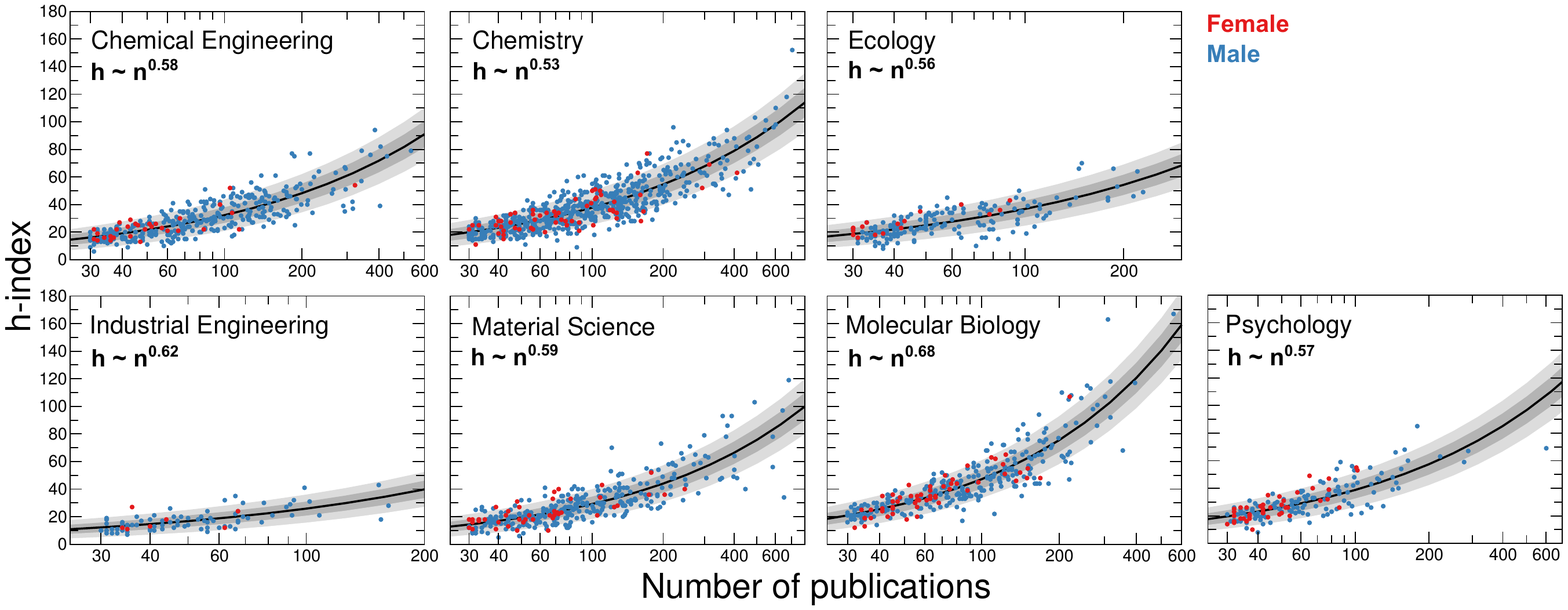}
\caption{ 
}
\label{fig7}
\end{center}
\end{figure}

\clearpage
\begin{figure}[!ht]
\begin{center}
\vskip .7cm
\includegraphics[width=\textwidth,clip=true,trim=0 0 0 0]{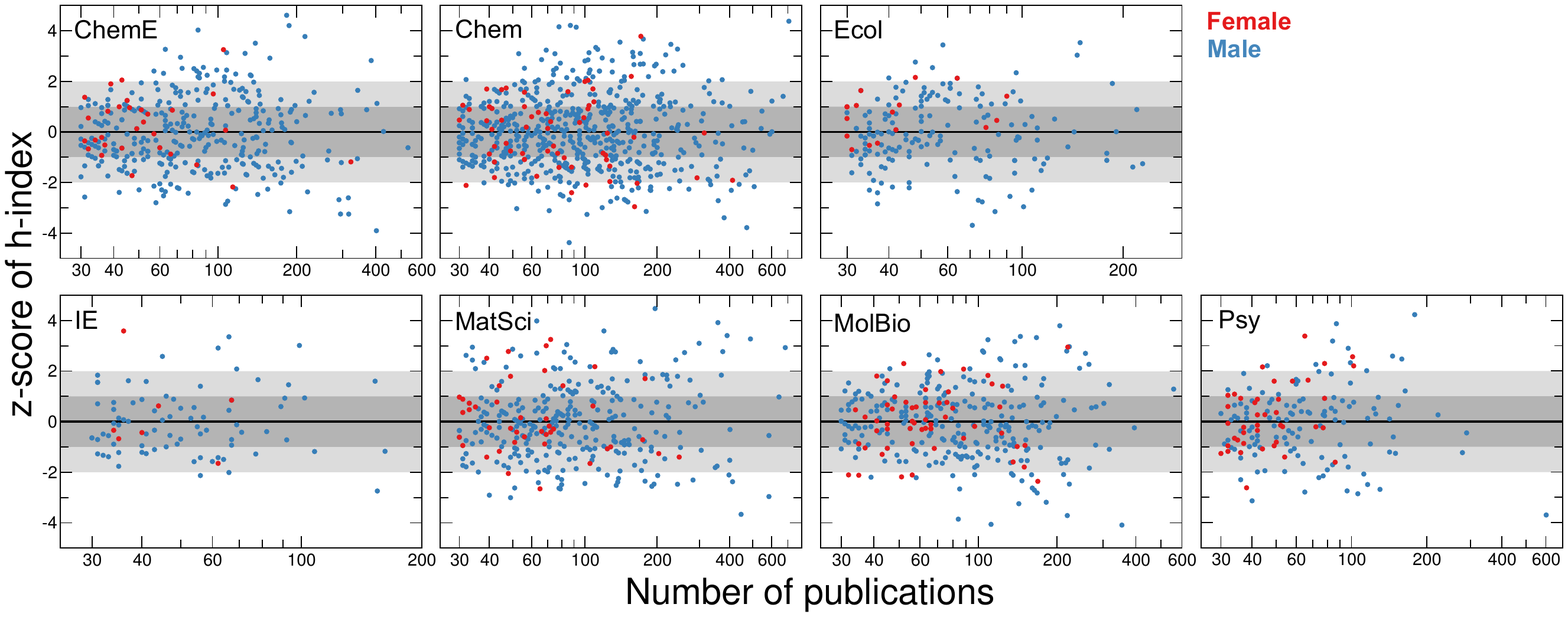}
\caption{ 
}
\label{fig8}
\end{center}
\end{figure}

\clearpage
\begin{figure}[!ht]
\begin{center}
\vskip .7cm
\includegraphics[width=\textwidth,clip=true,trim=0 0 0 0]{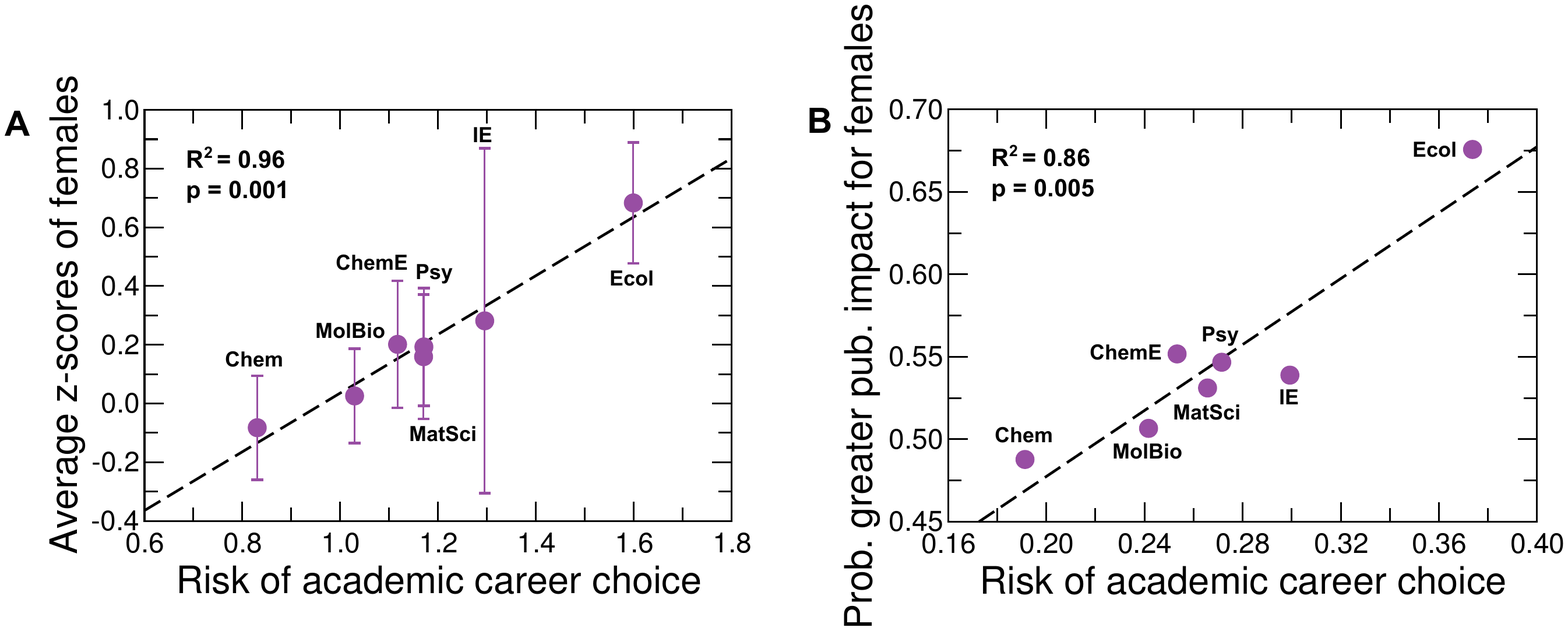}
\caption{ 
}
\label{fig9}
\end{center}
\end{figure}

\newpage
\clearpage
\section*{Supplementary Figure Legends}
\pagestyle{empty}

\paragraph*{Figure S1}
\textbf{Statistical significance of gender difference in publication rate.}
Probability that a female faculty member published more articles at a given stage of her career than a male peer at the same career stage (red lines). We use z-scores to account for two trends in the data: (i) the publication rate increases over years (Fig.~\ref{fig3}), and (ii) the publication rate varies with the career length (Fig.~\ref{fig4}). We indicate the 90\% and 95\% confidence intervals by the dark grey and light grey areas respectively, and the medians of the probabilities obtained from random ensembles by black lines.

\paragraph*{Figure S2}
\textbf{Time to career independence of female faculty members.} 
The fraction of publications authored by female faculty members in which the female faculty
  member is the last author (red diamonds) and the fraction of
  publications in which a faculty member is the first author (pink
  squares). The red/pink lines are fits of the data to a
  generalized logistic function (Methods, Table~S\ref{tables11}--S\ref{tables17}). The grey shaded areas
  indicate the periods of professional independence for the different
  disciplines.

\paragraph*{Figure S3}
\textbf{Time to career independence of male faculty members.} 
The fraction of publications authored by male faculty members in which the male faculty
  member is the last author (blue diamonds) and the fraction of
  publications in which a faculty member is the first author (azure
  squares). The blue/azure lines are fits of the data to a
  generalized logistic function (Methods, Table~S\ref{tables11}--S\ref{tables17}). The grey shaded areas
  indicate the periods of professional independence for the different
  disciplines.

\paragraph*{Figure S4}  
\textbf{Statistical significance of gender difference in publication impact.}
Probability that female authors have larger \emph{h}-index than male authors when accounting for the number of publications. The red line shows the results for windows including authors with at least 30 publications and at most $N_{\rm{max}}$ publications. Dark grey areas and light grey areas show the 90\% and 95\% confidence intervals (see Methods for details).


\newpage
\clearpage
\section*{Supporting Information}
\pagestyle{empty}

\paragraph*{Confidence interval for publication rates.}
In Fig.~S\ref{figs1}, we consider the probability that a female author has a publication rate higher than that of a male author as a function of the career stage. Since not all the authors have started their careers in the same year and the publication rate is increasing with time, we consider standard scores relative to career stages in stead of raw publication numbers.

In Fig.~S\ref{figs1}, we show for each career stage $t$ the quantity $P\left[z_F(t) > z_M(t) \right]$, representing
the probability that a female author has a standard score higher than that of a male author at the same stage of her career. We also compute the confidence intervals for these probability values, in the hypothesis that the difference
is not due to gender-related reasons. We generate the confidence intervals valid under this hypothesis using a re-sampling method: The populations of females and males are fixed, the values of all standard scores are also fixed, 
but values of the standard score are randomly reassigned among authors (this is the same as randomly reassigning the genders to authors). For each random configuration, we compute again the probability $P\left[z_F(t) > z_M(t) \right]$
and obtain the confidence intervals by repeating this procedure 1000 times.

\paragraph*{Confidence interval for publication impact.}
In Fig.~S\ref{figs4}, we consider the probability that a female author has higher publication impact than that of a male author as a function of the number of publications. The analysis is similar to that for publication rates in two aspects: (i) that we use standard scores instead of raw \emph{h}-indices, and (ii) the method we use to obtain the confidence intervals. The differences are (i) that we consider the \emph{h}-index and the number of publications instead of the number of publications and the career stage, respectively, and (ii) that the mean and standard deviation are given by Eq. \ref{eq4}.


\newpage
\clearpage

\renewcommand{\tablename}{Table S}
\setcounter{table}{0}

\begin{table}
\caption{{\bf Gender of faculty in Chemical Engineering departments.}}
\begin{center}
\begin{tabular}{lrr}
{\bf Department} & {\bf Male} & {\bf Female}\\\midrule
California Institute of Technology&7&3\\
Carnegie Mellon University&	19&	4\\
Cornell University&	16&	1\\
Georgia Institute of Technology&	31&	7\\
Johns Hopkins University&	11&	2\\
Massachusetts Institute of Technology&	35&	4\\
North Carolina State University&	22&	2\\
Northwestern University&	14&	3\\
Ohio State University&	14&	3\\
Pennsylvania State University&	18&	4\\
Princeton University&	19&	3\\
Purdue University&	23&	4\\
Rensselaer Polytechnic Institute&	16&	1\\
Rice University&	13&	5\\
Stanford University&	17&	3\\
University of California, Berkeley&	15&	3\\
University of California, Davis&	20&	7\\
University of California, Los Angeles&	11&	2\\
University of California, Santa Barbara&	17&	2\\
University of Colorado&	18&	4\\
University of Delaware&	20&	4\\
University of Florida&	20&	2\\
University of Illinois at Urbana Champaign&	17&	3\\
University of Massachusetts Amherst&	15&	3\\
University of Michigan&	18&	5\\
University of Minnesota at Minneapolis St. Paul&	30&	3\\
University of Notre Dame&	17&	3\\
University of Pennsylvania&	21&	2\\
University of Texas at Austin&	21&	2\\
University of Washington&	15&	2\\
University of Wisconsin at Madison&	17&	2\\\hline
\bf Total & \bf 567 & \bf 98\\
\end{tabular}
\label{tables1}
\end{center}
\end{table}

\newpage
\clearpage
\begin{table}
\caption{{\bf Gender of faculty in Chemistry departments.}}
\begin{center}
\begin{tabular}{lrr}
{\bf Department} & {\bf Male} & {\bf Female}\\\midrule
Boston University	&	20	&	3	\\
California Institute of Technology	&	21	&	5	\\
Carnegie Mellon University	&	24	&	5	\\
Cornell University	&	23	&	3	\\
Duke University	&	24	&	3	\\
Emory University	&	18	&	2	\\
Georgia Institute of Technology	&	38	&	4	\\
Harvard University	&	20	&	4	\\
Johns Hopkins University	&	20	&	1	\\
Massachusetts Institute of Technology	&	25	&	7	\\
North Carolina State University	&	22	&	6	\\
Northwestern University	&	26	&	4	\\
Ohio State University	&	33	&	8	\\
Pennsylvania State University 	&	30	&	6	\\
Princeton University 	&	17	&	3	\\
Purdue University 	&	40	&	15	\\
Rensselaer Polytechnic Institute 	&	19	&	3	\\
Rice University	&	19	&	3	\\
Stanford University	&	20	&	3	\\
University of California, Berkeley 	&	50	&	8	\\
University of California, Davis	&	28	&	10	\\
University of California, Los Angeles 	&	43	&	11	\\
University of California, Santa Barbara	&	36	&	4	\\
University of Colorado 	&	43	&	8	\\
University of Delaware 	&	25	&	7	\\
University of Florida 	&	38	&	6	\\
University of Illinois at Urbana Champaign 	&	37	&	7	\\
University of Massachusetts Amherst 	&	23	&	4	\\
University of Michigan 	&	36	&	13	\\
University of Minnesota at Minneapolis St. Paul	&	36	&	6	\\
University of Notre Dame 	&	30	&	5	\\
University of Pennsylvania 	&	28	&	6	\\
University of Texas at Austin 	&	41	&	5	\\
University of Washington 	&	31	&	4	\\
University of Wisconsin at Madison 	&	40	&	6	\\\hline
\bf Total & \bf 1,024 & \bf 198\\
\end{tabular}
\label{tables2}
\end{center}
\end{table}

\newpage
\clearpage
\begin{table}
\caption{{\bf Gender of faculty in Ecology departments.}}
\begin{center}
\begin{tabular}{lrr}
{\bf Department} & {\bf Male} & {\bf Female}\\\midrule
Cornell University&	19&	6\\
Duke University&	24&	7\\
Harvard University&	8&	3\\
Pennsylvania State University&	41&	17\\
Princeton University&	14&	2\\
Rice University&	9&	4\\
Stanford University&	12&	6\\
University of California, Berkeley&	33&	10\\
University of Chicago&	17&	3\\
University of Georgia&	20&	4\\
University of Illinois at Urbana Champaign&	54&	18\\
University of Michigan&	37&	14\\
University of Texas at Austin& 11&	5\\
University of Washington&	4&	2\\
University of Wisconsin at Madison&	26&	5\\\hline
\bf Total & \bf 329 & \bf 108\\
\end{tabular}
\label{tables3}
\end{center}
\end{table}
\clearpage

\newpage
\clearpage
\begin{table}
\caption{{\bf Gender of faculty in Industrial Engineering departments.}}
\begin{center}
\begin{tabular}{lrr}
{\bf Department} & {\bf Male} & {\bf Female}\\\midrule
Cornell University & 19 & 1\\
Georgia Institute of Technology & 47 & 10\\
North Carolina State University	& 19 & 1\\
Northwestern University & 17 & 1\\
Pennsylvania State University & 19 & 3\\
Purdue University & 17 &2\\
Stanford University & 27 & 7\\
University of California, Berkeley & 15 & 4\\
University of Florida & 14 & 1\\
University of Illinois at Urbana Champaign & 17 & 4\\
University of Michigan & 23 & 4\\
University of Minnesota at Minneapolis St. Paul & 6 & 2\\
University of Texas at Austin & 6 & 1\\
University of Washington & 4 & 4\\
University of Wisconsin at Madison & 12 & 6\\\hline
\bf Total & \bf 262 & \bf 51\\
\end{tabular}
\label{tables4}
\end{center}
\end{table}

\newpage
\clearpage
\begin{table}
\caption{{\bf Gender of faculty in Material Science departments.}}
\begin{center}
\begin{tabular}{lrr}
{\bf Department} & {\bf Male} & {\bf Female}\\\midrule
Boston University	&	23	&	8	\\
California Institute of Technology	&	8	&	3	\\
Carnegie Mellon University 	&	17	&	2	\\
Cornell University	&	14	&	3	\\
Duke University	&	21	&	4	\\
Georgia Institute of Technology	&	31	&	9	\\
Johns Hopkins University	&	10	&	2	\\
Massachusetts Institute of Technology	&	19	&	6	\\
North Carolina University	&	12	&	4	\\
Northwestern University	&	30	&	5	\\
Ohio State University	&	21	&	4	\\
Pennsylvania State University	&	20	&	4	\\
Purdue University	&	13	&	2	\\
Rensselaer Polytechnic Institute	&	9	&	2	\\
Rice University	&	14	&	1	\\
Stanford University	&	13	&	1	\\
University of California, Berkeley 	&	18	&	7	\\
University of California, Los Angeles	&	17	&	2	\\
University of California, Santa Barbara	&	31	&	3	\\
University of Delaware	&	12	&	2	\\
University of Florida	&	34	&	6	\\
University of Illinois at Urbana Champaign	&	19	&	4	\\
University of Michigan	&	26	&	5	\\
University of Pennsylvania	&	15	&	5	\\
University of Washington	&	12	&	3	\\
University of Wisconsin at Madison	&	16	&	4	\\\hline
\bf Total & \bf 475& \bf 101\\
\end{tabular}
\label{tables5}
\end{center}
\end{table}

\newpage
\clearpage
\begin{table}
\caption{{\bf Gender of faculty in Molecular Biology departments.}}
\begin{center}
\begin{tabular}{lrr}
{\bf Department} & {\bf Male} & {\bf Female}\\\midrule
Caltech	&	12	&	3	\\
Harvard University	&	34	&	10	\\
Johns Hopkins University	&	24	&	8	\\
MIT	&	46	&	21	\\
Princeton University	&	34	&	20	\\
Stanford University	&	14	&	10	\\
University of California, Berkeley	&	32	&	9	\\
University of California, San Francisco	&	39	&	11	\\
University of Texas at Austin	&	93	&	30	\\
Washington University in St. Louis	&	122	&	35	\\
Yale University	&	26	&	12	\\\hline
\bf Total & \bf 476 & \bf 169\\
\end{tabular}
\label{tables6}
\end{center}
\end{table}
\clearpage

\newpage
\clearpage
\begin{table}
\caption{{\bf Gender of faculty in Psychology departments.}}
\begin{center}
\begin{tabular}{lrr}
{\bf Department} & {\bf Male} & {\bf Female}\\\midrule
Harvard University	&	17	&	9	\\
Princeton University	&	19	&	11	\\
Stanford University	&	21	&	11	\\
University of California, Berkeley	&	19	&	12	\\
University of California, Los Angeles	&	41	&	26	\\
University of Illinois, Urbana Champaign	&	35	&	21	\\
University of Michigan	&	65	&	46	\\
University of Minnesota at Minneapolis	&	33	&	8	\\
University of Wisconsin at Madison	&	18	&	18	\\
Yale University	&	14	&	11	\\\hline
\bf Total & \bf 282& \bf 173\\
\end{tabular}
\label{tables7}
\end{center}
\end{table}


\newpage
\clearpage
\begin{table}
\caption{{\bf Estimated values of parameters of logistic function for Chemical Engineering data.}}
\begin{center}
\begin{tabular}{ccccccc}
\toprule
\multirow{2}{*}{\bf Discipline}
 & \multirow{2}{*}{\bf Gender} & \multirow{2}{*}{\bf Authorship} & \multicolumn{4}{c}{\bf Parameter estimates}\\
 & & & $A$ & $K$ & $B$ & $M$ \\ \midrule
\multirow{6}{*}{Chemical Engineering} 
	& \multirow{2}{*}{All}
		& First & 0.62 $\pm$ 0.01 & 0.08 $\pm$0.00 & 0.64 $\pm$ 0.03 & 4.9 $\pm$ 0.1\\ 
		&& Last & 0.09 $\pm$ 0.01 & 0.60 $\pm$0.01 & 0.80 $\pm$ 0.07 & 5.4 $\pm$ 0.1\\ \cline{2-7}
	& \multirow{2}{*}{Female}
		& First & 0.71 $\pm$ 0.06 & 0.00 $\pm$ 0.01 & 0.50 $\pm$ 0.08 & 4.1 $\pm$ 0.4\\ 
		&& Last & 0.01 $\pm$ 0.04 & 0.63 $\pm$ 0.02 & 0.6 $\pm$ 0.1 & 5.0 $\pm$ 0.3\\ \cline{2-7}
	& \multirow{2}{*}{Male}
		& First & 0.61 $\pm$ 0.01 & 0.09 $\pm$0.00 & 0.67 $\pm$ 0.04 & 5.0 $\pm$ 0.1\\ 
		&& Last & 0.10 $\pm$ 0.01 & 0.59 $\pm$0.00 & 0.85 $\pm$ 0.04 & 5.4 $\pm$ 0.1\\
\bottomrule
\end{tabular}
\label{tables11}
\end{center}
\end{table}

\newpage
\clearpage
\begin{table}
\caption{{\bf Estimated values of parameters of logistic function for Chemistry data.}}
\begin{center}
\begin{tabular}{ccccccc}
\toprule
\multirow{2}{*}{\bf Discipline}
 & \multirow{2}{*}{\bf Gender} & \multirow{2}{*}{\bf Authorship} & \multicolumn{4}{c}{\bf Parameter estimates}\\
 & & & $A$ & $K$ & $B$ & $M$ \\ \midrule
\multirow{6}{*}{Chemistry} 
	& \multirow{2}{*}{All}
		& First & 0.43 $\pm$ 0.01 & 0.13 $\pm$0.01 & 0.64 $\pm$ 0.07 & 5.9 $\pm$ 0.2\\ 
		&& Last & 0.20 $\pm$ 0.01 & 0.60 $\pm$0.01 & 0.63 $\pm$ 0.08 & 6.2 $\pm$ 0.2\\ \cline{2-7}
	& \multirow{2}{*}{Female}
		& First & 0.46 $\pm$ 0.01 & 0.09 $\pm$ 0.01 & 0.67 $\pm$ 0.09 & 5.8 $\pm$ 0.2\\ 
		&& Last & 0.14 $\pm$ 0.02 & 0.61 $\pm$ 0.01 & 0.7 $\pm$ 0.1 & 6.3 $\pm$ 0.2\\ \cline{2-7}
	& \multirow{2}{*}{Male}
		& First & 0.43 $\pm$ 0.01 & 0.13 $\pm$0.01 & 0.63 $\pm$ 0.08 & 5.9 $\pm$ 0.2\\ 
		&& Last & 0.21 $\pm$ 0.02 & 0.60 $\pm$0.01 & 0.63 $\pm$ 0.09 & 6.2 $\pm$ 0.2\\
\bottomrule
\end{tabular}
\label{tables12}
\end{center}
\end{table}

\newpage
\clearpage
\begin{table}
\caption{{\bf Estimated values of parameters of logistic function for Ecology data.}}
\begin{center}
\begin{tabular}{ccccccc}
\toprule
\multirow{2}{*}{\bf Discipline}
 & \multirow{2}{*}{\bf Gender} & \multirow{2}{*}{\bf Authorship} & \multicolumn{4}{c}{\bf Parameter estimates}\\
 & & & $A$ & $K$ & $B$ & $M$ \\ \midrule
\multirow{6}{*}{Ecology}
	& \multirow{2}{*}{All}
		& First & 0.41 $\pm$ 0.01 & 0.18 $\pm$ 0.02 & 0.5 $\pm$ 0.1 & 8.3 $\pm$ 0.5\\ 
		&& Last & 0.14 $\pm$ 0.01 & 0.44 $\pm$ 0.02 & 0.5 $\pm$ 0.1 & 8.2 $\pm$ 0.3\\ \cline{2-7}
	& \multirow{2}{*}{Female}
		& First & 0.41 $\pm$ 0.03 & 0.16 $\pm$ 0.03 & 0.5 $\pm$ 0.3 & 8 $\pm$ 1\\ 
		&& Last & 0.12 $\pm$ 0.03 & 0.49 $\pm$ 0.02 & 0.6 $\pm$ 0.2 & 8.5 $\pm$ 0.6\\ \cline{2-7}
	& \multirow{2}{*}{Male}
		& First & 0.41 $\pm$ 0.02 & 0.18 $\pm$ 0.02 & 0.5 $\pm$ 0.1 & 8.4 $\pm$ 0.6\\ 
		&& Last & 0.15 $\pm$ 0.01 & 0.43 $\pm$ 0.02 & 0.4 $\pm$ 0.1 & 8.1 $\pm$ 0.4\\
\bottomrule
\end{tabular}
\label{tables13}
\end{center}
\end{table}

\newpage
\clearpage
\begin{table}
\caption{{\bf Estimated values of parameters of logistic function for Industrial Engineering data.}}
\begin{center}
\begin{tabular}{ccccccc}
\toprule
\multirow{2}{*}{\bf Discipline}
 & \multirow{2}{*}{\bf Gender} & \multirow{2}{*}{\bf Authorship} & \multicolumn{4}{c}{\bf Parameter estimates}\\
 & & & $A$ & $K$ & $B$ & $M$ \\ \midrule
\multirow{6}{*}{Industrial Engineering}
	& \multirow{2}{*}{All}
		& First & 0.37 $\pm$ 0.04 & 0.18 $\pm$ 0.05 & 0.3 $\pm$ 0.2 & 8 $\pm$ 2\\ 
		&& Last & 0.21 $\pm$ 0.02 & 0.45 $\pm$ 0.01 & 0.4 $\pm$ 0.1 & 6.1 $\pm$ 0.6\\ \cline{2-7}
	& \multirow{2}{*}{Female}
		& First & 0.39 $\pm$ 0.03 & 0.20 $\pm$ 0.04 & 1 $\pm$ 1 & 9 $\pm$ 1\\ 
		&& Last & 0.23 $\pm$ 0.06 & 0.44 $\pm$ 0.04 & 0.6 $\pm$ 0.6 & 7 $\pm$ 2\\ \cline{2-7}
	& \multirow{2}{*}{Male}
		& First & 0.4 $\pm$ 0.2 & 0.1 $\pm$ 0.1 & 0.2 $\pm$ 0.3 & 6 $\pm$ 4\\ 
		&& Last & 0.20 $\pm$ 0.05 & 0.45 $\pm$ 0.02 & 0.4 $\pm$ 0.2 & 6 $\pm$ 1\\
\bottomrule
\end{tabular}
\label{tables14}
\end{center}
\end{table}

\newpage
\clearpage
\begin{table}
\caption{{\bf Estimated values of parameters of logistic function for Material Science data.}}
\begin{center}
\begin{tabular}{ccccccc}
\toprule
\multirow{2}{*}{\bf Discipline}
 & \multirow{2}{*}{\bf Gender} & \multirow{2}{*}{\bf Authorship} & \multicolumn{4}{c}{\bf Parameter estimates}\\
 & & & $A$ & $K$ & $B$ & $M$ \\ \midrule
\multirow{6}{*}{Material Science}
	& \multirow{2}{*}{All}
		& First & 0.43 $\pm$ 0.02 & 0.08 $\pm$ 0.01 & 0.37 $\pm$ 0.04 & 6.0 $\pm$ 0.3\\ 
		&& Last & 0.17 $\pm$ 0.01 & 0.46 $\pm$ 0.01 & 0.48 $\pm$ 0.06 & 6.6 $\pm$ 0.3\\ \cline{2-7}
	& \multirow{2}{*}{Female}
		& First & 0.7 $\pm$ 0.5 & 0.0 $\pm$ 0.1 & 0.2 $\pm$ 0.2 & 2 $\pm$ 8\\ 
		&& Last & 0.10 $\pm$ 0.05 & 0.56 $\pm$ 0.05 & 0.3 $\pm$ 0.1 & 7.7 $\pm$ 0.8\\ \cline{2-7}
	& \multirow{2}{*}{Male}
		& First & 0.42 $\pm$ 0.01 & 0.09 $\pm$ 0.01 & 0.41 $\pm$ 0.04 & 6.2 $\pm$ 0.2\\ 
		&& Last & 0.18 $\pm$ 0.01 & 0.45 $\pm$ 0.01 & 0.52 $\pm$ 0.07 & 6.5 $\pm$ 0.3\\
\bottomrule
\end{tabular}
\label{tables15}
\end{center}
\end{table}

\newpage
\clearpage
\begin{table}
\caption{{\bf Estimated values of parameters of logistic function for Molecular Biology data.}}
\begin{center}
\begin{tabular}{ccccccc}
\toprule
\multirow{2}{*}{\bf Discipline}
 & \multirow{2}{*}{\bf Gender} & \multirow{2}{*}{\bf Authorship} & \multicolumn{4}{c}{\bf Parameter estimates}\\
 & & & $A$ & $K$ & $B$ & $M$ \\ \midrule
\multirow{6}{*}{Molecular Biology}
	& \multirow{2}{*}{All}
		& First & 0.42 $\pm$ 0.01 & 0.11 $\pm$ 0.01 & 0.54 $\pm$ 0.08 & 7.2 $\pm$ 0.3\\ 
		&& Last & 0.16 $\pm$ 0.01 & 0.57 $\pm$ 0.01 & 0.48 $\pm$ 0.04 & 7.3 $\pm$ 0.2\\ \cline{2-7}
	& \multirow{2}{*}{Female}
		& First & 0.42 $\pm$ 0.03 & 0.09 $\pm$ 0.03 & 0.5 $\pm$ 0.2 & 7.6 $\pm$ 0.7\\ 
		&& Last & 0.14 $\pm$ 0.03 & 0.65 $\pm$ 0.03 & 0.42 $\pm$ 0.09 & 7.8 $\pm$ 0.5\\ \cline{2-7}
	& \multirow{2}{*}{Male}
		& First & 0.42 $\pm$ 0.01 & 0.11 $\pm$ 0.01 & 0.54 $\pm$ 0.07 & 7.1 $\pm$ 0.2\\ 
		&& Last & 0.17 $\pm$ 0.01 & 0.56 $\pm$ 0.01 & 0.49 $\pm$ 0.04 & 7.3 $\pm$ 0.2\\
\bottomrule
\end{tabular}
\label{tables16}
\end{center}
\end{table}

\newpage
\clearpage
\begin{table}
\caption{{\bf Estimated values of parameters of logistic function for Psychology data.}}
\begin{center}
\begin{tabular}{ccccccc}
\toprule
\multirow{2}{*}{\bf Discipline}
 & \multirow{2}{*}{\bf Gender} & \multirow{2}{*}{\bf Authorship} & \multicolumn{4}{c}{\bf Parameter estimates}\\
 & & & $A$ & $K$ & $B$ & $M$ \\ \midrule
\multirow{6}{*}{Psychology}
	& \multirow{2}{*}{All}
		& First & 0.41 $\pm$ 0.01 & 0.26 $\pm$ 0.01 & 0.9 $\pm$ 0.3 & 8.5 $\pm$ 0.5\\ 
		&& Last & 0.18 $\pm$ 0.01 & 0.38 $\pm$ 0.01 & 0.7 $\pm$ 0.1 & 8.2 $\pm$ 0.3\\ \cline{2-7}
	& \multirow{2}{*}{Female}
		& First & 0.43 $\pm$ 0.02 & 0.30 $\pm$ 0.02 & 1.8 $\pm$ 2.2 & 8.5 $\pm$ 0.7\\ 
		&& Last & 0.19 $\pm$ 0.02 & 0.33 $\pm$ 0.02 & 1.1 $\pm$ 0.8 & 8.0 $\pm$ 0.7\\ \cline{2-7}
	& \multirow{2}{*}{Male}
		& First & 0.41 $\pm$ 0.01 & 0.25 $\pm$ 0.01 & 0.7 $\pm$ 0.3 & 8.3 $\pm$ 0.6\\ 
		&& Last & 0.17 $\pm$ 0.01 & 0.40 $\pm$ 0.02 & 0.54 $\pm$ 0.15 & 8.1 $\pm$ 0.5\\
\bottomrule
\end{tabular}
\label{tables17}
\end{center}
\end{table}

\newpage
\clearpage
\begin{table}
\caption{{\bf Estimated values of parameters of the power law relation between impact and number of publications, $\mathbf{h=kn^{\alpha}}$.}}
\begin{center}
\begin{tabular}{lrr}
\toprule
{\bf Discipline} & $\alpha$ & k\\\midrule
Chemical Engineering & 0.58 & 2.3\\
Chemistry & 0.53 & 3.3\\
Ecology & 0.56 & 2.8\\
Industrial Engineering & 0.62 & 1.5\\
Material Science & 0.59 & 1.9\\
Molecular Biology & 0.68 & 2.1\\
Psychology & 0.57 & 2.8\\
\bottomrule
\end{tabular}
\label{tables18}
\end{center}
\end{table}

\clearpage
\newpage
\renewcommand{\figurename}{Figure S}
\setcounter{figure}{0}

\clearpage
\begin{figure}[!ht]
\begin{center}
\vskip .7cm
\includegraphics[width=\textwidth,clip=true,trim=0 180 0 110]{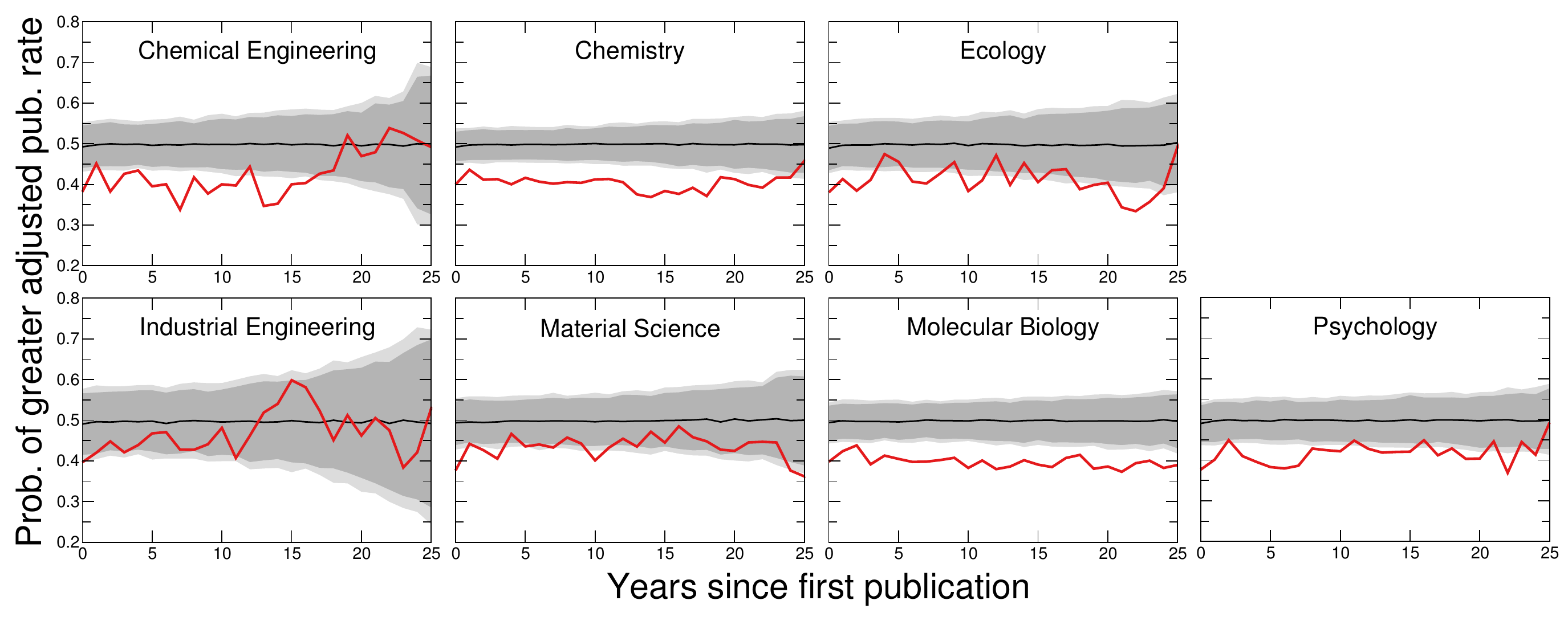}
\caption{ 
}
\vskip .1cm
\label{figs1}
\end{center}
\end{figure}

\clearpage
\begin{figure}[!ht]
\begin{center}
\vskip .7cm
\includegraphics[width=\textwidth,clip=true,trim=0 180 0 120]{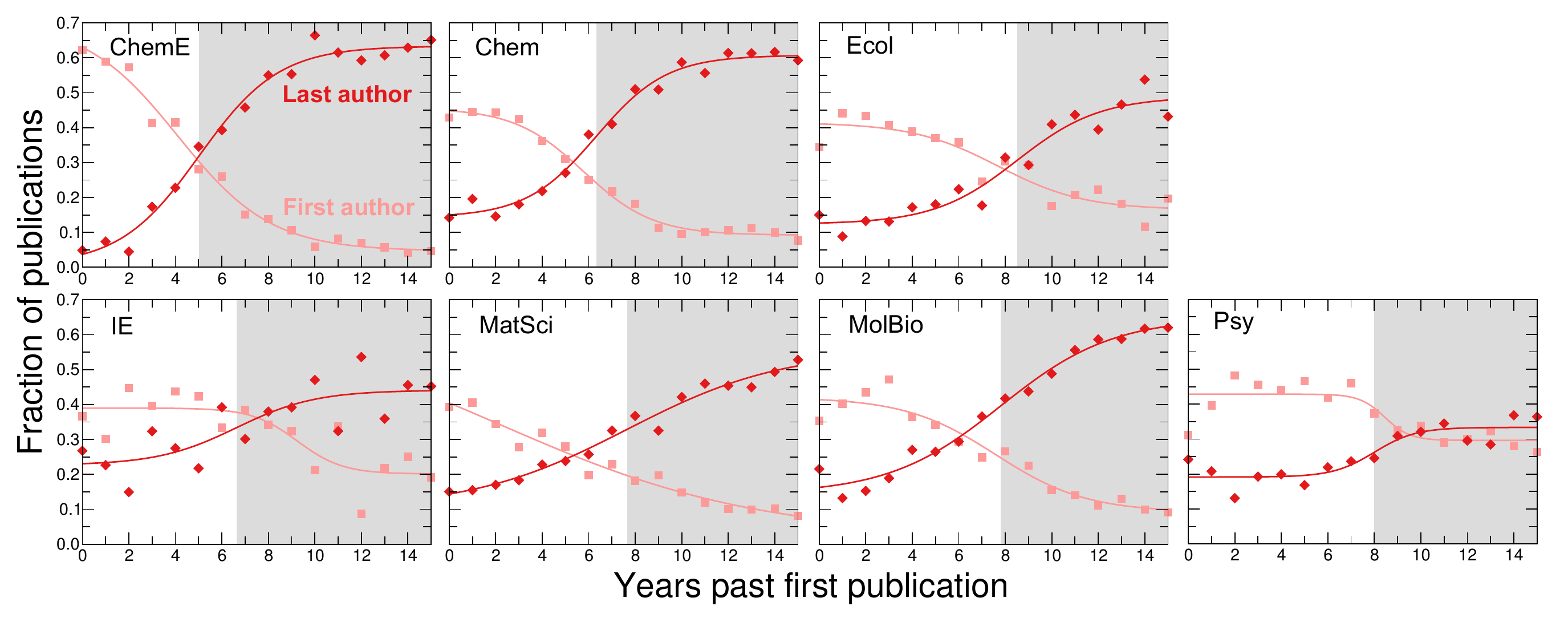}
\caption{
}
\vskip .1cm
\label{figs2}
\end{center}
\end{figure}

\clearpage
\begin{figure}[!ht]
\begin{center}
\vskip .7cm
\includegraphics[width=\textwidth,clip=true,trim=0 200 0 100]{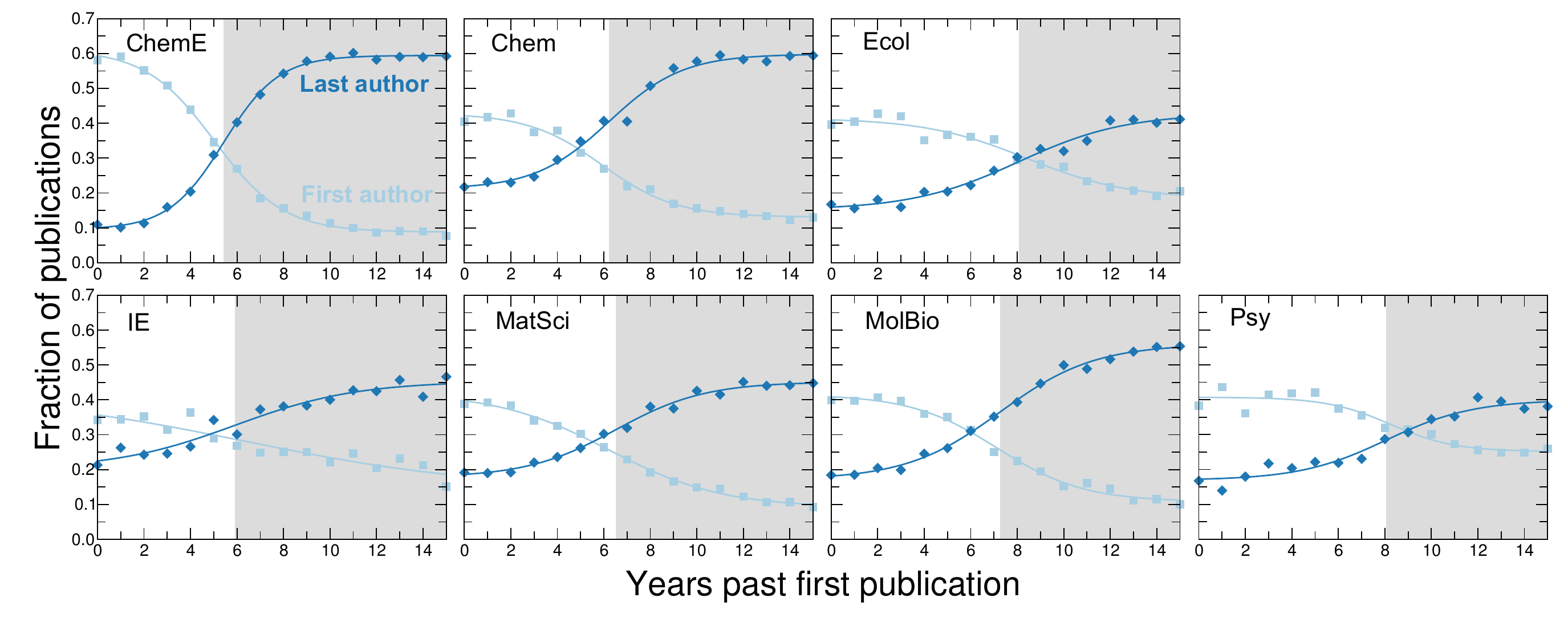}
\caption{
}
\vskip .1cm
\label{figs3}
\end{center}
\end{figure}

\clearpage
\begin{figure}[!ht]
\begin{center}
\vskip .7cm
\includegraphics[width=\textwidth,clip=true,trim=0 170 0 110]{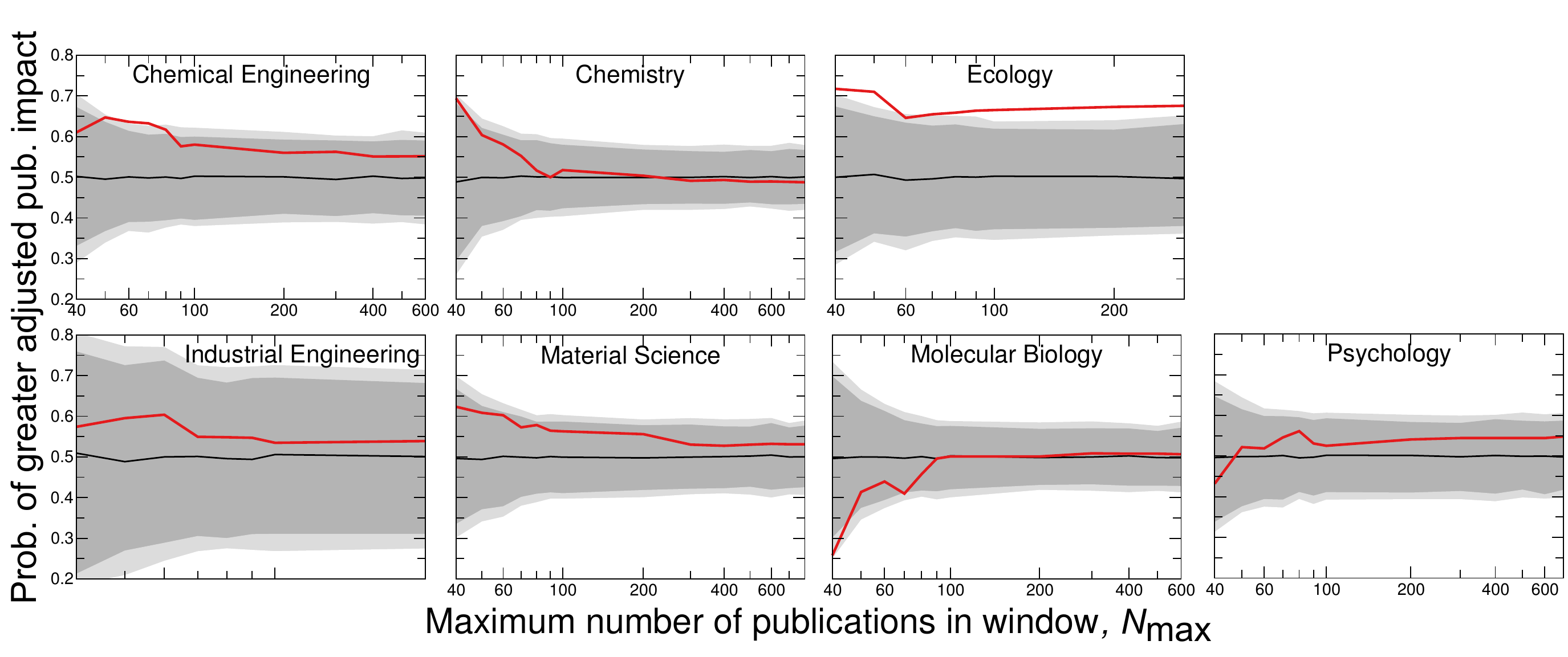}
\caption{
}
\vskip .1cm
\label{figs4}
\end{center}
\end{figure}


\end{document}